\documentclass[longauth]{aa}  
\pdfoutput=1
\usepackage{times}
\usepackage[T1]{fontenc}
\usepackage{pslatex}
\usepackage{amsmath}
\usepackage{color,graphicx}
\usepackage{rotating}
\usepackage{txfonts}
\usepackage{float}
\usepackage[colorlinks=true, citecolor=red]{hyperref}
\begin{document} 

\title{\textit{H}-band discovery of additional Second-Generation stars in the Galactic Bulge Globular Cluster NGC 6522 as observed by APOGEE and \textit{Gaia}} 	
 
\author{
 Jos\'e G. Fern\'andez-Trincado\inst{1, 2, 3},
 	  O. Zamora\inst{4, 5},    
 	  Diogo Souto\inst{6, 7},  
 	  R. E. Cohen\inst{8}, 
 	  F. Dell'Agli\inst{4, 5},	 
 	  D. A. Garc\'ia-Hern\'andez\inst{4, 5}, 
 	  T. Masseron\inst{4, 5}, 
 	  R. P. Schiavon\inst{9}, 
 	  Sz. M\'esz\'aros\inst{10, 11}, 
 	  K. Cunha\inst{12, 6},   
 	  Sten Hasselquist\inst{13}, 
 	  M. Shetrone\inst{14}, 
 	  J. Schiappacasse Ulloa\inst{3},  
 	  B. Tang\inst{15},  	  
 	  D. Geisler\inst{3, 16, 17},	 
 	  D. R. G. Schleicher\inst{3}, 
 	  S. Villanova\inst{3},   
 	  R. E. Mennickent\inst{3},  
 	  D. Minniti\inst{18, 19, 20}, 
 	  J. Alonso-Garc\'ia\inst{21, 19}, 
 	  A. Manchado\inst{4, 5, 22},  
 	  T. C. Beers\inst{23},  
 	  J. Sobeck\inst{24}, 	    
 	  G. Zasowski\inst{25}, 
 	  M. Schultheis\inst{26},  
 	  S. R. Majewski\inst{27}, 
 	  A. Rojas-Arriagada\inst{28}, 	   
 	  A. Almeida\inst{16}, 
  	  F. Santana\inst{29},
 	  R. J. Oelkers\inst{30},
 	  P. Longa-Pe\~na\inst{21},
 	  R. Carrera\inst{4, 5},  
 	  A. J. Burgasser\inst{31}, 
 	  R. R. Lane\inst{28},	 
 	  A. Roman-Lopes\inst{16},
 	  Inese I. Ivans\inst{25}
 	  \& F. R. Hearty\inst{32}
 }

 \institute{
 	     Instituto de Astronom\'ia y Ciencias Planetarias, Universidad de Atacama, Copayapu 485, Copiap\'o, Chile
 	     \and
	   	 Institut Utinam, CNRS UMR 6213, Universit\'e Bourgogne-Franche-Comt\'e, OSU THETA Franche-Comt\'e, Observatoire de Besan\c{c}on, BP 1615, 25010 Besan\c{c}on Cedex, France  
	   	 \and
	   	 Departamento de Astronom\'\i a, Casilla 160-C, Universidad de Concepci\'on, Concepci\'on, Chile   	 
	   	 \and
	   	 Instituto de Astrof\'{\i}sica de Canarias, V\'{\i}a L\'actea S/N, E-38205 La Laguna, Tenerife, Spain
	   	 \and
	   	 Departamento de Astrof\'{\i}sica, Universidad de La Laguna (ULL), E-38206 La Laguna, Tenerife, Spain
	   	 \and   	 
	   	 Observat\'orio Nacional, Rua Gal. Jos\'e Cristino 77, 20921-400 Rio de Janeiro, Brazil
	   	 \and
	   	 Departamento de F\'isica, Universidade Federal de Sergipe, Av. Marechal Rondon, S/N, 49000-000 S\~ao Crist\'ov\~ao, SE, Brazil   
	   	 \and
	   	 Space Telescope Science Institute, 3700 San Martin Dr, Baltimore, MD 21218, USA
	   	 \and
	   	 Astrophysics Research Institute, Liverpool John Moores University, 146 Brownlow Hill, Liverpool, L3 5RF, United Kingdom
	   	 \and
	   	 ELTE Gothard Astrophysical Observatory, H-9704 Szombathely, Szent Imre Herceg st. 112, Hungary
	   	 \and
	   	 Premium Postdoctoral Fellow of the Hungarian Academy of Sciences
	   	 \and 
	   	 Steward Observatory, University of Arizona, 933 North Cherry Avenue, Tucson, AZ 85721, USA
	   	 \and
	   	 New Mexico State University, Las Cruces, NM 88003, USA
	   	 \and
	   	 University of Texas at Austin, McDonald Observatory, Fort Davis, TX 79734, USA
	   	 \and	   	 
	   	 School of Physics and Astronomy, Sun Yat-sen University, Zhuhai 519082, China
	   	 \and
	   	 Departamento de F\'isica, Facultad de Ciencias, Universidad de La Serena, Cisternas 1200, La Serena, Chile
	     \and 
	     Instituto de Investigación Multidisciplinario en Ciencia y Tecnología, Universidad de La Serena. Avenida Raúl Bitrán S/N, La Serena, Chile	 	 
	    \and	 
	   	 Departamento de Fisica, Facultad de Ciencias Exactas, Universidad Andres Bello, Av. Fernandez Concha 700, Las Condes, Santiago, Chile
	   	 \and
	   	 Instituto Milenio de Astrofisica, Santiago, Chile
	   	 \and
	   	 Vatican Observatory, V00120 Vatican City State, Italy
	   	 \and
	   	 Centro de Astronom\'{i}a (CITEVA), Universidad de Antofagasta, Av. Angamos 601, Antofagasta, Chile
	   	 \and
	   	 Consejo Superior de Investigaciones Cient\'ificas, Spain
	   	 \and
	   	 Department of Physics and JINA Center for the Evolution of the Elements, University of Notre Dame, Notre Dame, IN 46556, USA
	   	 \and  
	   	 Department of Astronomy, University of Washington, Seattle, WA, 98195, USA
	   	 \and
	   	 Department of Physics \& Astronomy, University of Utah, Salt Lake City, UT, 84112, USA
	   	 \and
	   	 Laboratoire Lagrange, Universit\'e C\^ote d'Azur, Observatoire de la C\^ote d'Azur, CNRS, Blvd de l'Observatoire, F-06304 Nice, France
	   	 \and
	   	 Department of Astronomy, University of Virginia, Charlottesville, VA 22903, USA
	   	 \and
	   	 Instituto de Astrof\'sica, Pontificia Universidad Cat\'olica de Chile, Av. Vicuna Mackenna 4860, 782-0436 Macul, Santiago, Chile
	   	 \and
	   	 Universidad de Chile, Av. Libertador Bernardo O'Higgins 1058, Santiago de Chile
	   	 \and
	   	 Department of Physics \& Astronomy, Vanderbilt University, Nashville, TN, 37235, USA
	   	 \and
	   	 Department of Physics, University of California, San Diego, CA 92093, USA
	   	 \and
	   	 Department of Astronomy and Astrophysics, The Pennsylvania State University, University Park, PA 16802   
 }

 \date{Received XX/XX/2018; Accepted XX/XX/2018}
 \titlerunning{NGC 6522}

\authorrunning{J. G. Fern\'andez-Trincado et al.} 

 
  \abstract
   {We present elemental abundance analysis of high-resolution spectra for five giant stars, deriving Fe, Mg, Al, C, N, O, Si and Ce abundances, and spatially located within the innermost regions of the bulge globular cluster NGC 6522, based on \textit{H}-band spectra taken with the multi-object APOGEE-north spectrograph from the SDSS-IV Apache Point Observatory Galactic Evolution Experiment (APOGEE) survey. { Of the five cluster candidates, two previously unremarked stars are confirmed to have \textit{second-generation} (SG) abundance patterns, with the basic pattern of depletion in C and Mg simultaneous with enrichment in N and Al as seen in other SG globular cluster populations at similar metallicity. } In agreement with the most recent optical studies, the NGC 6522 stars analyzed exhibit (when available) only mild overabundances of the \textit{s}-process element Ce, contradicting the idea of the NGC 6522 stars being formed from gas enriched by spinstars and indicating that other stellar sources such as massive AGB stars could be the primary intra-cluster medium polluters. The peculiar abundance signature of SG stars have been observed in our data, confirming the presence of multiple generations of stars in NGC 6522.}     
   \maketitle
  
  \section{Introduction}
  \label{section1}
 
  The presence of multiple populations (MPs) with distinctive light-element abundances were recently identified in several bulge globular clusters \citep[see, ][for instance]{Schiavon2017a, Recio-Blanco2017, Tang2017, Cesar2017}. In particular, \citet{Schiavon2017a} have studied the chemical composition of a few red giant stars within the bulge globular clusters (GCs) NGC 6553, NGC 6528, Terzan 5, Palomar 6, and NGC 6522, using near infrared (1.5-1.7$\mu$m) high-resolution (\textit{R}$=$22,000) APOGEE spectra from the 12th data release \citep[DR12,][]{Alam2015}. These studies have also included the re-reduced and re-calibrated spectra of the latest APOGEE DR13\footnote{\textit{APOGEE field -- BAADEWIN\_001-04:} Particularly in this field, APOGEE/DR13/DR14 have the same targets as APOGEE DR12, but the data reduction and calibration have been improved in several ways. For more details we refer the reader to a forthcoming paper (Holtzman et al. in preparation).} 
  data release \citep{Albareti2017} for the globular cluster NGC 6553 \citep[e.g.,][]{Tang2017}, where we have included more chemical species with reliable light-element abundances (namely O, Na, Si, Ca, Cr, Mn, and Ni). \citet{Schiavon2017a} and \citet{Tang2017} have provided useful chemical "tags" in several elemental abundances for several Milky Way bulge globular cluster stars with clear signatures of \textit{polluted chemistry}; i.e., they have found the distinctive chemical patterns characterising multiple populations, with comparable chemical behavior to what is reported in extensive spectroscopic and photometry survey of GCs in general \citep[see][]{Gratton2004, Gratton2007, Carretta2007, Carretta2009a, Carretta2009b, Carretta2010, Gratton2012, Meszaros2015, Garcia-Hernandez2015, Carretta2016, Recio-Blanco2017, Pancino2017, Schiavon2017a, Meszaros2017, Tang2017, Bastian2018, Tang2018, Kerber2018}. 
  
  Large-scale spectroscopic surveys like APOGEE \citep[see][]{Majewski2017} have confirmed that several bulge GCs exhibit significant star-to-star abundance variations in their light-element content \citep[see][]{Schiavon2017a, Tang2017}, with the usual anti-correlations between pairs of light elements, such as C-N and Al-N. This behaviour demonstrates that the CNO, NeNa, and MgAl cycles took place in these GCs \citep[see, e.g.,][]{Meszaros2015, Schiavon2017a, Tang2017, Pancino2017,  Ventura2016a, Flavia2017}.
  
  Following this line of investigation, we turn our attention to the low-mass \citep[$\sim$5.93$\times 10^{4}$ M$_{\odot}$:][]{Gnedin1997} and old \citep[$\sim$ 12.5 and 13.0 Gyr:][]{Kerber2018} bulge globular cluster NGC 6522. Earlier studies show that this ancient Milky Way globular cluster hosts remarkably high abundances of slow neutron-capture (\textit{s}-process) elements \citep[e.g.,][]{Chiappini2011}. \citet[][]{Chiappini2011} interpreted this observation as evidence of NGC 6522 stars being formed from gas enriched by massive fast-rotating stars \citep[spinstars; see][]{Pignatari2008}, which possibly makes NGC 6522 distinct from other GCs. 
  
  However, more recent chemical re-analysis by \citet{Ness2014}  and \citet{Barbuy2014} found no enhancement in the \textit{s}-process elements for the same stars previously studied by \citet[][]{Chiappini2011}. ). The abundances they find can be explained by mass transfer from \textit{s}-process-rich asymptotic giant branch (AGB)
  stars or alternative self-enrichment scenarios (e.g., the massive AGBs
  self-enrichment scenario) without invoking massive fast-rotating stars. \citet{Kerber2018}, based on detailed analysis of HST proper-motion-cleaned color-magnitude diagrams, found that NGC 6522 exhibits at least two stellar populations with an intrinsically wide subgiant branch, consistent with a first and second stellar generation.
  
  Here we carry out a detailed re-analysis of the NGC 6522 field to search for abundance anomalies through the line-by-line spectrum synthesis calculations for the full set of (atomic and molecular) lines (particularly CN, OH, CO, Al, Mg, and Si) in the re-reduced APOGEE DR14 spectra \citep{Abolfathi2017}. The phenomenon of star-to-star light-element abundance variations in NGC 6522 indicates the presence of \textit{multiple stellar populations}, such as those claimed by \citet{Schiavon2017a} and \citet{Recio-Blanco2017}, and provides crucial observational evidence that NGC 6522 could be the fossil relic of one of the structures that contributed to generate the N-rich population towards the Milky Way bulge \citep{Schiavon2017b}. It also reinforces the link between GCs and the chemical anomalies (\textit{second-generation field stars}\footnote{Here, the term \textit{second-generation} refers to groups of stars in globular clusters that display altered light-element abundances (C, N, O, Na, Al, and Mg).}) recently found toward the Galactic bulge field \citep[e.g.,][]{Fernandez-Trincado2017a, Fernandez-Trincado2019c}, as well as that with the N-rich moderately metal-poor halo stars \citep{Martell2010, Martell2011, Martell2016, Tang2019}, mimicking the chemical abundance patterns of the \textit{second-generation} population of globular clusters \citep[see][]{Fernandez-Trincado2016b, Fernandez-Trincado2019a, Fernandez-Trincado2019b, Fernandez-Trincado2019c, Fernandez-Trincado2019d}. { More recently, observations extending the analysis to other elements have already detected departures from what seemed to be a simple chemical evolutionary path, like the existence of a Na-rich population toward the outer bulge likely originated from disrupted GCs \citep[e.g,][]{Lee2019}.}
  
  This article is structured as follows.  We describe the data in Section \ref{section2}. We describe the cluster membership selection in Section \ref{section3}. In Section \ref{section4} and \ref{neutron} we provide our abundance analysis for light and heavy elements, respectively. In Section \ref{results} we discuss the results. We present our conclusions in Section \ref{section5}.

  \section{APOGEE DATA}
  \label{section2}
  
  High-resolution (R$\sim$22,500) H-band spectroscopic ($\lambda=$ 1.51 - 1.69$\mu$m) observations were obtained with the Apache Point Observatory Galactic Evolution Experiment (APOGEE), as part of Sloan Digital Sky Survey IV that observed 277,000 stars in the Milky Way \citep[see][]{Gunn2006, Eisenstein2011, Wilson2012, Majewski2017}. Here we use the most recent re-reduced and re-calibrated APOGEE spectra from the 14th data release of SDSS \citep[DR14,][]{Abolfathi2017}.
  
  We have re-analyzed available APOGEE spectra towards the Baade's window (APOGEE field: BAADEWIN\_001-04) region around (l, b)$\approx$(1$^{\circ}$, -4$^{\circ}$) with a field of view of $\sim$ 3 sq. degree, comprising 460 stars \citep[for details, see][]{Zasowski2013, Zasowski2017}.
  
  One of our stars in the BAADEWIN\_001-04 field, 2M18032356-3001588, was recently studied in \citet[][]{Schiavon2017a} using the DR12 datasets throught ASPCAP\footnote{ASPCAP: The APOGEE Stellar Parameter and Chemical Abundances Pipeline} results \citep{GarciaPerez2016a}. The same authors have suggested the presence of MPs in NGC 6522 based on the polluted chemistry (high Al and N) observed in 2M18032356-3001588, this hypothesis has been recently supported by similar analysis from the Gaia-ESO survey \citep[see][]{Recio-Blanco2017}. Here we present an independent analysis using the newly released APOGEE DR14 stellar spectra towards NGC 6522, and report the identification of four new potential cluster members with polluted chemistry towards the innermost regions of the cluster. 
  
  It is to be noted here that the new highest likelihood cluster members (4 stars) were originally missed by \citet{Schiavon2017a}, because they adopted more rigorous limits on the NGC 6522 parameter space (radial velocity, metallicity, T$_{\rm eff}$, etc) as well as higher restrictions on the signal-to-noise (S/N) ratio ($>$ 70 pixel$^{-1}$) of the APOGEE spectra. In the next section, we present our adopted softer limits that take into account the updated parameter space of NGC 6522 and that have allowed us to identify new potential cluster members based on APOGEE data.
  
  \begin{figure}
  	\begin{center}
  		\includegraphics[width=95mm]{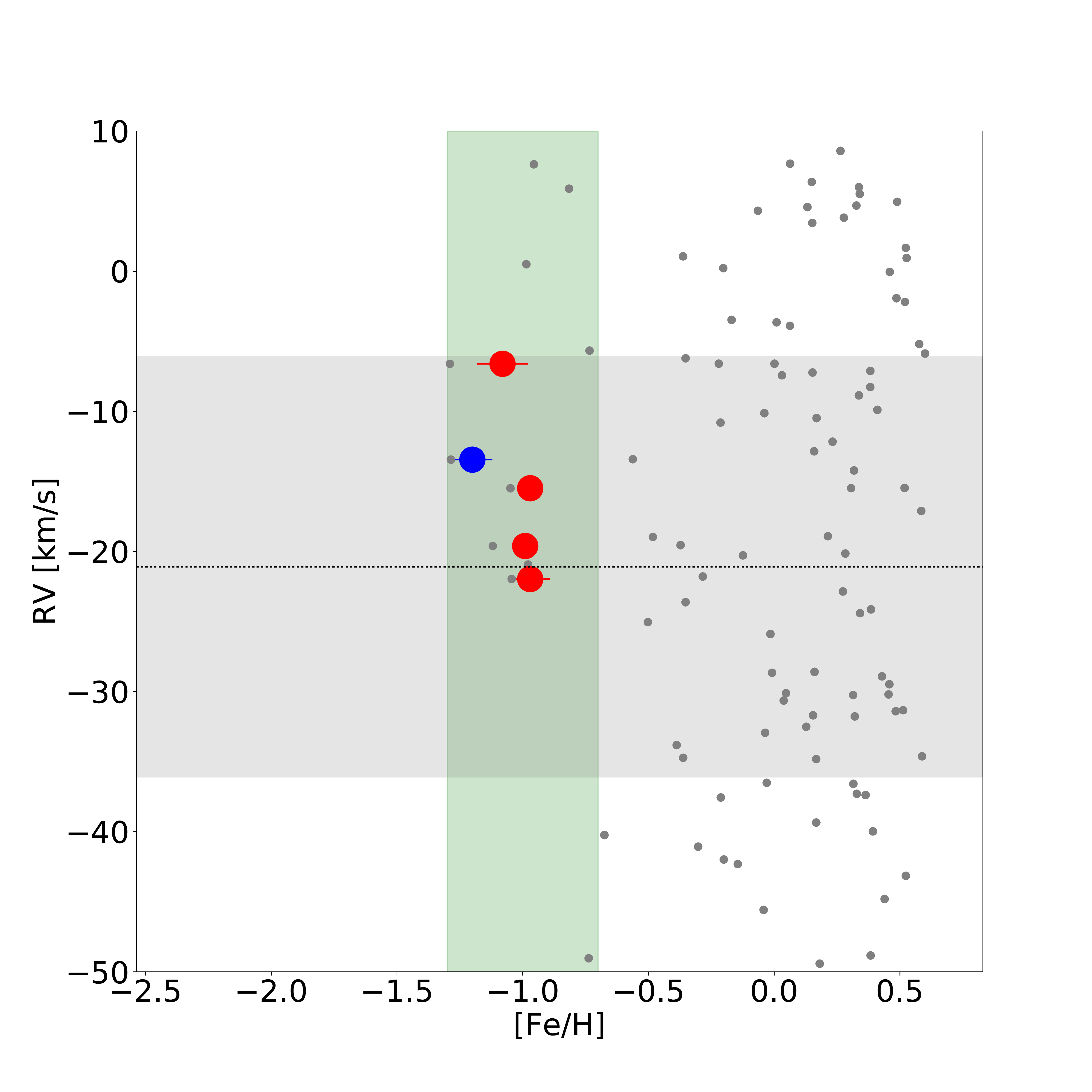}
  		\caption{The APOGEE/DR14 radial velocity of the stars against their metallicity (grey dots). The red filled circles are the new highest likelihood cluster members analyzed in this work, while the blue filled circle is the giant star analyzed in \citet{Schiavon2017a}. The grey and green shadow region defines the upper/lower limit for the membership selection, and the black dotted line marks the radial velocity (-21.1 km s$^{-1}$) of NGC 6522, according to \citet{Harris1996}.}
  		\label{FigureRV}
  	\end{center}
  \end{figure}
  
  \begin{figure}
  	\begin{center}
  		\includegraphics[width=85mm]{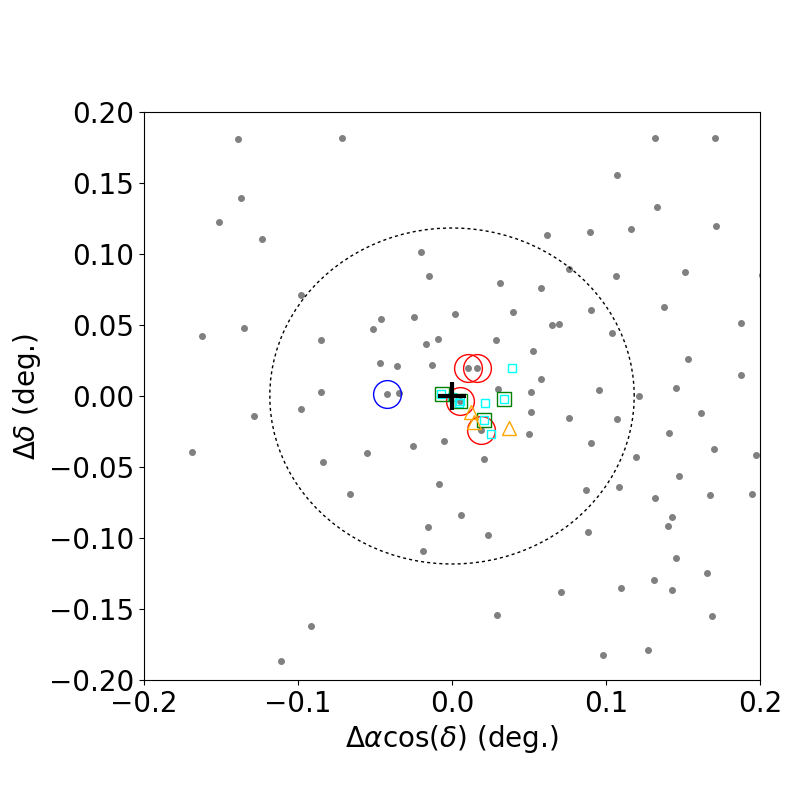}
  		\caption{Spatial distribution of targets in NGC 6522: Member candidates are highlighted with red open circles. The inner plus symbol is the centre of the cluster and the black dotted line marks the tidal radius of the cluster, $r_{\rm t} = 7.1^{+6.1}_{-3.7}$ arcmin. Field stars from the APOGEE survey located in the commissioning plate 4332, FIELD $=$ BAADEWIN\_001-04 are plotted using small gray symbols. The unfilled cyan squares, green squares, orange triangles and the blue open circle shows cluster members analysed in \citet{Ness2014},  \citet{Barbuy2014}, \citet{Recio-Blanco2017}, and \citet[][]{Schiavon2017a}, respectively.}
  		\label{Figure1}
  	\end{center}
  \end{figure}

    \begin{figure*}
    	\begin{center}
    		\includegraphics[width=200mm]{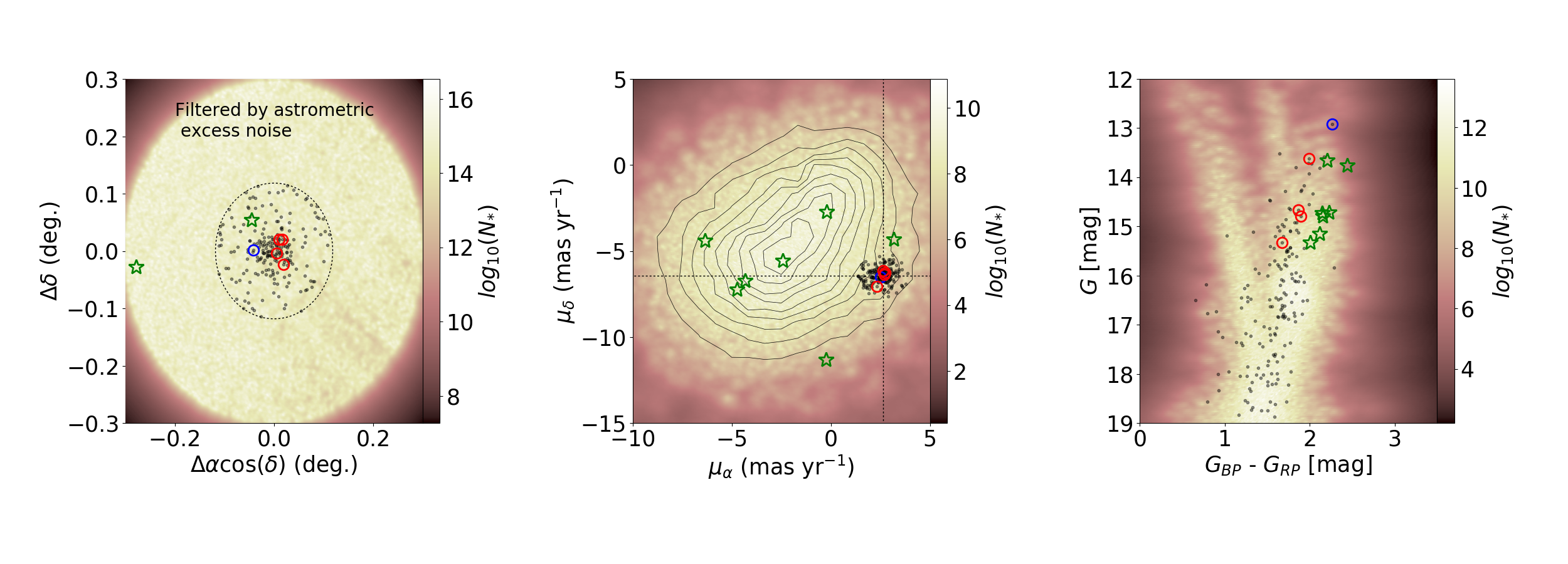}
    		\caption{ Kernel Density Estimate (KDE) smoothed distribution (\textit{left}), proper motion distribution (\textit{middle}) and color-magnitude diagram (\textit{right}) of Gaia DR2 stars toward the NGC 6522 field. Leftmost panel show the position of the newly detected member cluster candidates (red unfilled circles), the blue unfilled circle represent the star previously reported in \citet{Schiavon2017a}, the green unfilled stars show the position of the field stars contained within the green shadow region as illustrated in Figure \ref{FigureRV}, the black dots represent the Gaia DR2 stars inside the tidal radius of the cluster and whose proper motions match with the nominal proper motion of NGC 6522 within 3$\sigma_{\mu}$, whilst the black dotted circle mark the size of the tidal radius of cluster. Middle plot show ($\mu_{\alpha}$,$\mu_{\delta}$) distribution whilst the black dotted lines show the nominal proper motion of NGC 6522. Rightmost plot show the Gaia DR2 colour-magnitude diagram of each star.}
    		\label{Figure2a}
    	\end{center}
    \end{figure*}

  \section{CLUSTER MEMBERSHIP SELECTION}
  \label{section3}
  
  We selected probable cluster members based on the revised version of the structural parameters of NGC 6522, i.e., the cluster center ($\alpha$, $\delta$) $=$ (270.896805$^{\circ}$, -30.034204$^{\circ}$), with an uncertainty of 0.3 arcsec, from ellipse fitting to density maps from HST PSF photometry, and the tidal radius of the cluster, $r_t < 7.1^{+6.1}_{-3.7}$ arcmin. For a more detailed discussion, we refer the readers to a forthcoming paper (Cohen et al. 2018, in preparation). 
  
  To select the highest likelihood cluster members we also adopt a radial velocity range of $\langle RV \rangle \sim -21.1\pm 15$ km s$^{-1}$ \citep{Harris1996}. We have adopted a metallicity range of [Fe/H] $\sim -1.0 \pm 0.3$ dex \citep[e.g.,][]{Barbuy2009, Barbuy2014}; our stars are also recovered even adopting the cluster metallicity as reported in \citet{Ness2014}, [Fe/H]$= -1.15$. The radial velocity and metallicity of our stellar sample have been displayed in Figure \ref{FigureRV}, which shows that most stars have radial velocities and metallicities very close to the mean cluster values.
  
  In Figure \ref{Figure1}, we plot the spatial distribution of four new potential cluster members (2M18033819$-$3000515, 2M18033965$-$3000521, 2M18034052$-$3003281, and  2M18033660$-$3002164) against one star previously identified on APOGEE (2M18032356$-$3001588), which clearly lie near the cluster centre (all our candidate members fall within a relatively small radius, $\sim 2.5$ arcmin), as illustrated in the same figure. It is important to note that a detailed chemical analysis has not been done so far for these objects, except for: (i) 2M18033660$-$3002164, which was analyzed in \citet{Chiappini2011} and \citet{Ness2014} from GIRAFFE/VLT spectra. Unfortunately, this is the fainter star in our sample (see Table \ref{table1}) and its low-S/N APOGEE spectrum does not permit us to carry out a reliable and conclusive abundance analysis, especially for Al I lines; (ii) 2M18032356$-$3001588, was already studied by \citet[light elements: ][]{Schiavon2017a} and \citet[heavy elements:][]{Cunha2017}. We note, however, that we carry out an independent chemical analysis of 2M18032356-3001588 \citep[][]{Schiavon2017a}, which permit us to revisit its chemical composition.

  \begin{figure}
  	\begin{center}
  		\includegraphics[width=80mm]{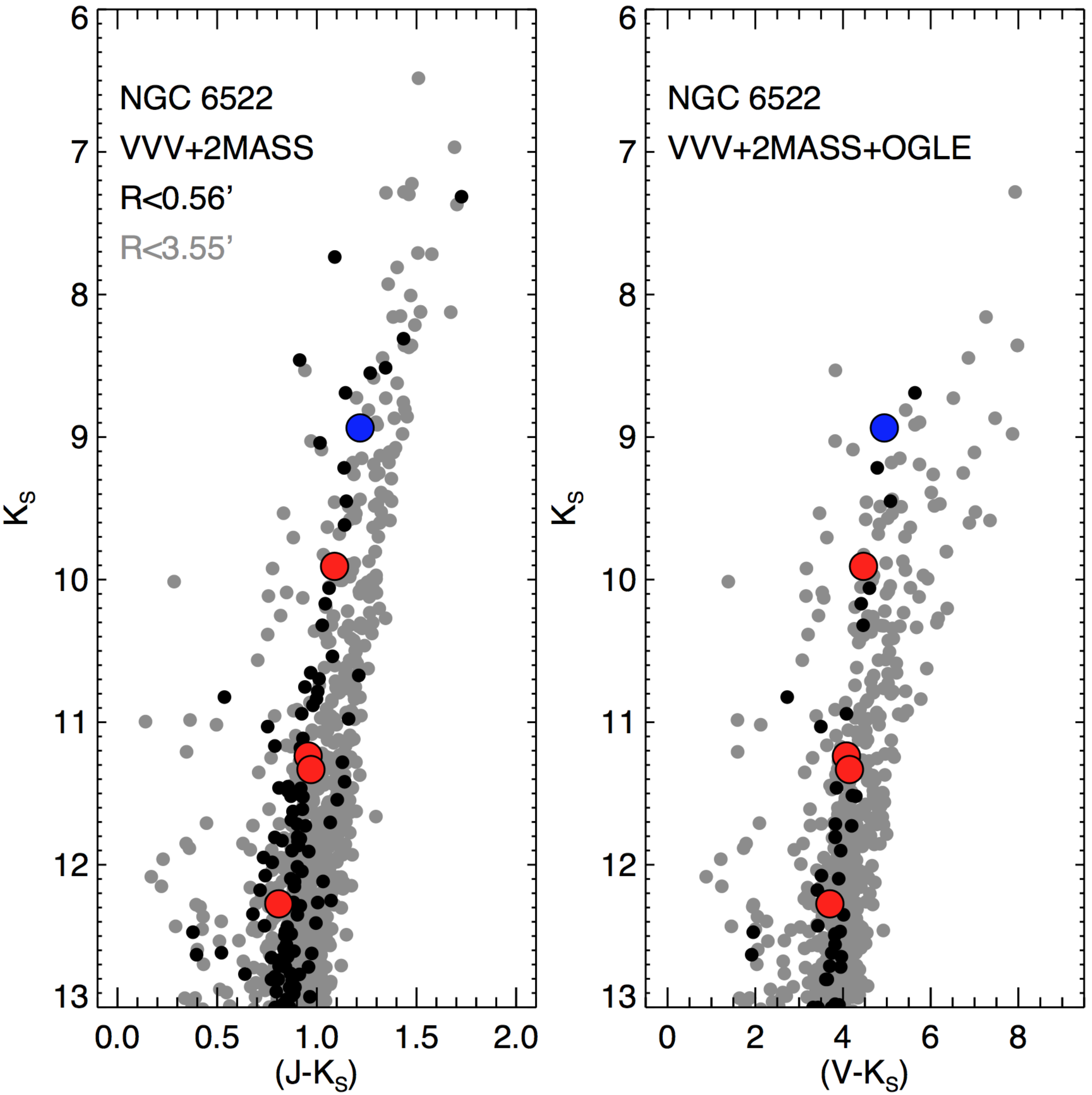}
  		\caption{Colour-Magnitude diagram of (VVV+2MASS K$_{\rm s}$, J- K$_{\rm s}$) and VVV+2MASS K$_{\rm s}$+OGLE from \citet{Cohen2017a}, illustrating the position all stars inside half the tidal radius (grey dots), and all inside the half-light radius (black dots), $r_{hl}=0.56^{+0.41}_{-0.12}$ arcmin with superimposed the position of the APOGEE spectroscopic targets. The symbols are the same as in Figure \ref{FigureRV}}
  		\label{Figure2}
  	\end{center}
  \end{figure}

  It is instructive to contrast the potential cluster candidate stars with those for NGC 6522 in the Gaia DR2 database \citep{gaiadr2}. Since NGC 6522 is relatively far ($d_{\odot} \sim 7.7$ kpc), we decided to pay particular attention to avoid contamination by data processing artifacts and/or spurious measurements. Therefore, we adopted the following conservative cuts on the columns of the Gaia DR2 \texttt{GAIA\_SOURCE} catalogue:

	\begin{itemize}
		\item[(1.)]   \texttt{ASTROMETRIC\_GOF\_AL $<$ 8}. This cut ensures that the statistics astrometric model resulted in a good fit to the data;
		\item[(2.)]  \texttt{ASTROMETRIC\_EXCESS\_NOISE\_SIG $\leq$ 2.} This criterion ensured that the selected stars were astrometrically well-behaved sources;
		\item[(3.)] \texttt{$-$0.23 $\leq$ MEAN\_VARPI\_FACTOR\_AL $\leq$ 0.32 AND} \texttt{VISIBILITY\_PERIODS\_USED > 7}. These cuts were used to exclude stars with parallaxes more vulnerable to errors;
		\item[(4.)] \texttt{G $<$ 19 mag}. This criterion minimized the chance of foreground contamination.
	\end{itemize}
	
The final sample so selected amounts to a total of 45,683 stars, which lie in a radius of 0.3 degree around the NGC 6522. Figure \ref{Figure2a} shows the spatial distribution, proper motion distribution and colour-magnitude diagram of the Gaia DR2 stars labeled as members (black dots) of NGC 6522 as well as Gaia DR2 field stars and the newly identified second-generation stars (blue and red unfilled circles). To select Gaia DR2 stars as potential members, we adopt $\sigma_{\mu}$ as the total uncertainty in quadrature obtained from a 2-dimensional Gaussian fit. For this purpose, a 2-dimensional Gaussian smoothing routine was applied in proper motion space for stars with $G<19$ mag within the { cluster tidal radius}. A 2D Gaussian was fitted to this sample and membership probabilities are assigned. With this procedure, we found: $\mu^{2D}_{\alpha} \pm \sigma_{\alpha} = 2.539 \pm 0.510$ mas yr$^{-1}$ and $\mu^{2D}_{\delta} \pm \sigma_{\delta} = -6.399 \pm 0.449$ mas yr$^{-1}$, and $\sigma_{\mu} = 0.608$ mas yr$^{-1}$, our results also agree remarkably well with the more recent { measurements of PMs for NGC 6522, e.g.: $\mu_{\alpha} = 2.618 \pm 0.072$ mas yr$^{-1}$, and $\mu_{\delta} = -6.431 \pm 0.071$ from \citet{Vasiliev18}.} A star was considered to be a GC member if its proper motion differs from that of NGC 6522 by not more than 3$\sigma_{\mu}$. One can see that the newly identified N-rich stars from the APOGEE survey are distributed along inside the tidal radius of the cluster and the proper motions of those stars match the nominal proper motion of NGC 6522, and the Gaia DR2 colour-magnitude diagram contains the stars with highest [N/Fe] in our sample along the red giant branch of NGC 6522. Based on the Gaia DR2 ($\mu_{\alpha}$,$\mu_{\delta}$) space, we rule out other possible cluster candidates in our APOGEE sample, which are highlighted by green unfilled symbols in Figure \ref{Figure2a} and lie in the green shadow region (grey dots) in Figure \ref{FigureRV}. 

    The position on the color-magnitude diagram (CMD) of the likely candidate members of NGC 6522 analyzed in this paper are shown in Figure \ref{Figure2}. One can immediately notice that the selected stars from the APOGEE survey lie in the upper part of the red giant branch (RGB) indicated by red and blue filled symbols, and occupy the same locus that other potential stellar cluster candidates inside the half-light radius $r_{hl}=0.56^{+0.41}_{-0.12}$ arcmin ---see \citet{Cohen2017a} for details about VVV+2MASS \citep{Skrutskie2006, Minniti2010} CMDs of this cluster. The faintest star in these diagrams correspond to the star 2M18033660$-$3002164 ($Ks\_{VVV} = 12.272$ and $G = 15.322$), while 2M18032356$-$3001588 ($Ks\_{VVV} = 9.157$ and $G = 12.920$) is the brightest star as listed in Table \ref{table1}.\\

  \section{LIGHT-ELEMENT abundances in NGC 6522}
  \label{section4}
  
  In this work, we employed the Brussels Automatic Stellar Parameter (BACCHUS)\footnote{The previous (DR12) and current
  	(DR13/14) version of ASPCAP does not determine the abundances of the
  	neutron-capture elements Ce and Nd, but the recent characterization (e.g.,
  	oscillator strenghts) of the H-band Nd II and Ce II lines \citep{Hasselquist2016, Cunha2017}, permits, in principle, their abundances derivation by
  	using a spectral synthesis code like BACCHUS (see text for more details). For
  	consistency (among other reasons), we re-derived all abundances with BACCHUS.} code \citep[see][]{Masseron2016, Hawkins2016} to derive chemical abundances for up to eight chemical elements that are typical indicators of stars with "polluted chemistry" in GCs { (C, N, O, Al, Mg, and Si) }\citep[see, e.g.,][]{Tang2017, Schiavon2017a}. The synthetic spectra were based on 1D Local Thermodynamic Equilibrium (LTE) model atmospheres calculated with  MARCS \citep{Gustafsson2008} using the solar abundance table from \citet{Asplund2005}, except for Ce, which we have adopted the abundance table from \citet[][]{Grevesse2015}. 
  
\begin{figure}
  	\begin{center}
  		\includegraphics[width=95mm]{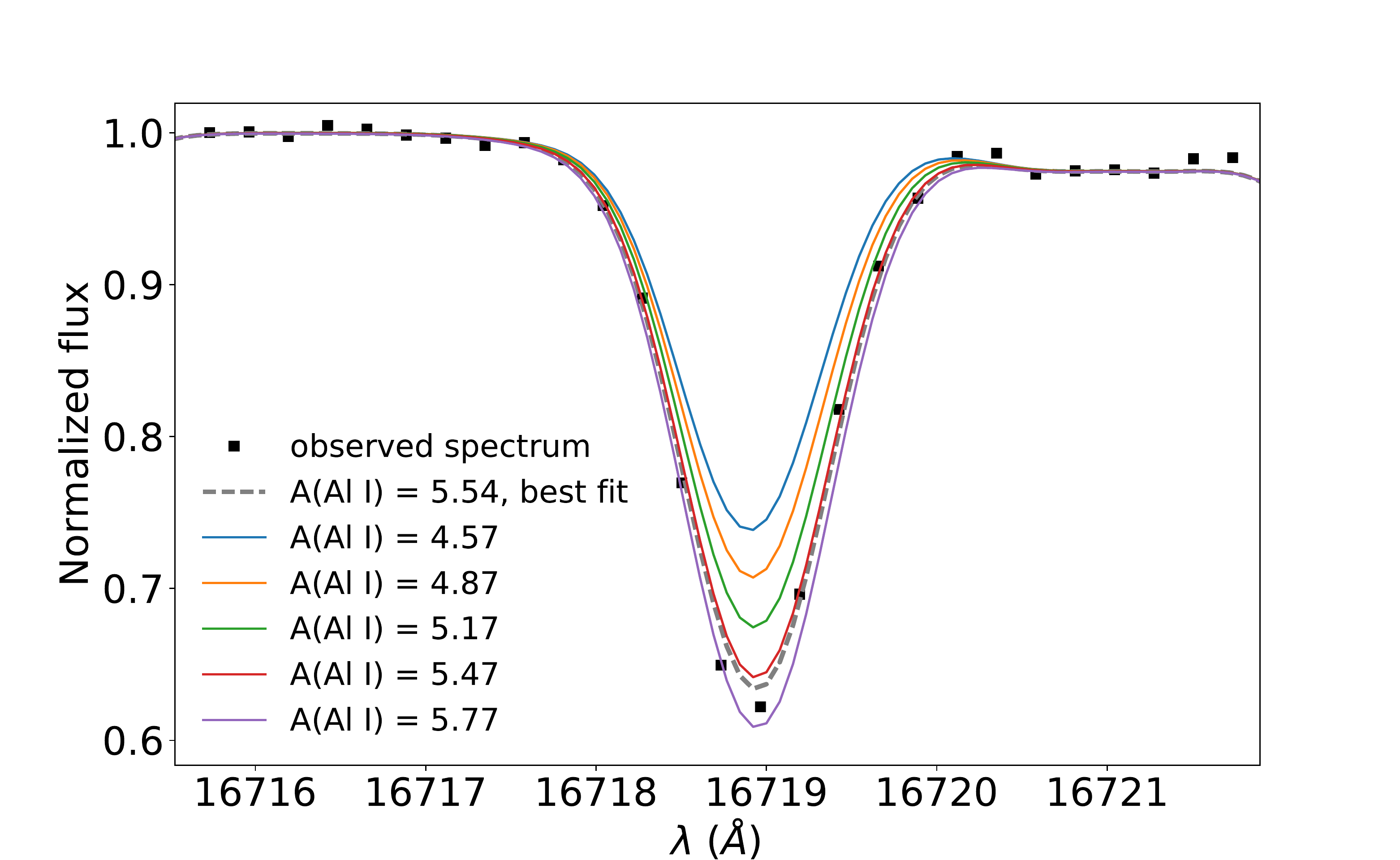}
  		\caption{High-resolution H--band observed spectrum of 2M18034052$-$3003281 (filled squares) in the 16716 -- 16721 \AA{} (Al I line) region. Superimposed are MARCS/BACCHUS synthetic spectra. All spectra are expressed in air wavelengths.}
  		\label{Figure4}
  	\end{center}
  \end{figure}

  All the chemical species were first visually inspected line-by-line and rejected if they were found to be problematic, i.e., lines heavily blended by telluric features were rejected. Note that, in contrast to ASPCAP pipeline \citep[which employ KURUCZ atmospheric models, e.g., see ][]{GarciaPerez2016a}, we provide a line-by-line analysis based on MARCS model atmosphere grid. Table \ref{table3} lists the wavelength regions used to obtain the individual abundances, while Figure \ref{Figure4} plots an example of the best fits obtained using MARCS/BACCHUS synthetic spectra around the Al I line, $\lambda_{\rm air} = $16718.957 \AA{}. BACCHUS software provides four different abundance determinations: (i) line-profile fitting; (ii) core line intensity comparison; (iii) global goodness-of-fit estimate ($\chi^2$); and (iv) equivalent width comparison; and each diagnostic yields validation flags. Based on these flags, a decision tree then rejects the line or accept it, keeping the best-fit abundance \citep[see, e.g.,][]{Hawkins2016}. Then, following the suggestion by \citet[][]{Hawkins2016}, we adopt the $\chi^{2}$ diagnostic as the abundance, which is the most robust.
  
\begin{figure*}[!ht]
  	\begin{center}
  		\includegraphics[width=160mm]{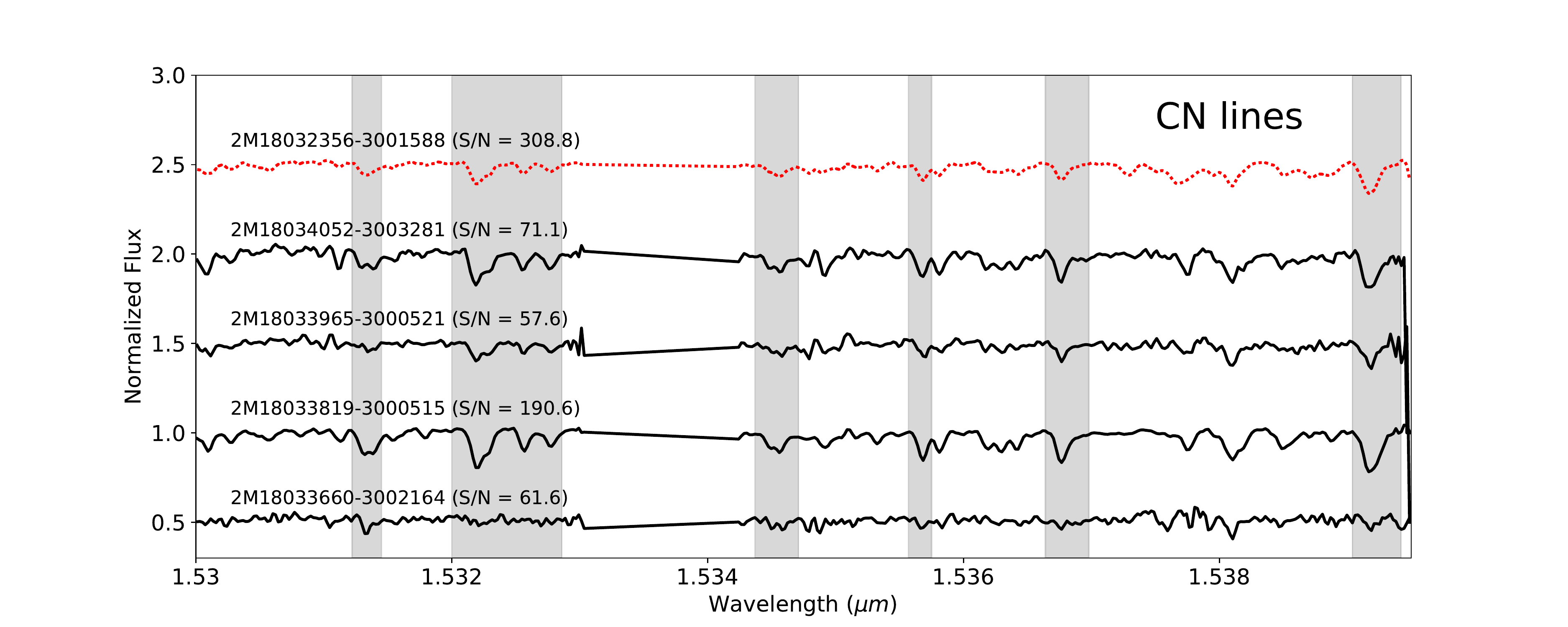}
  		\includegraphics[width=160mm]{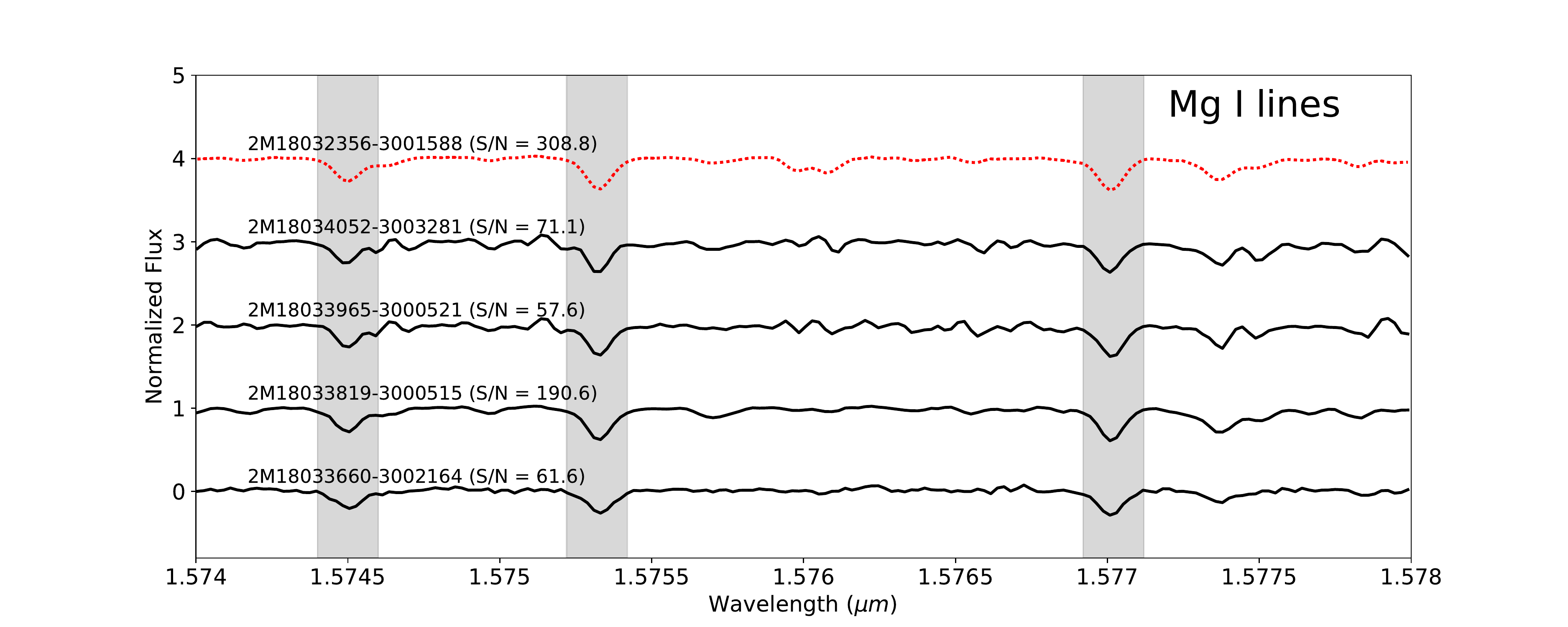}
  		\includegraphics[width=160mm]{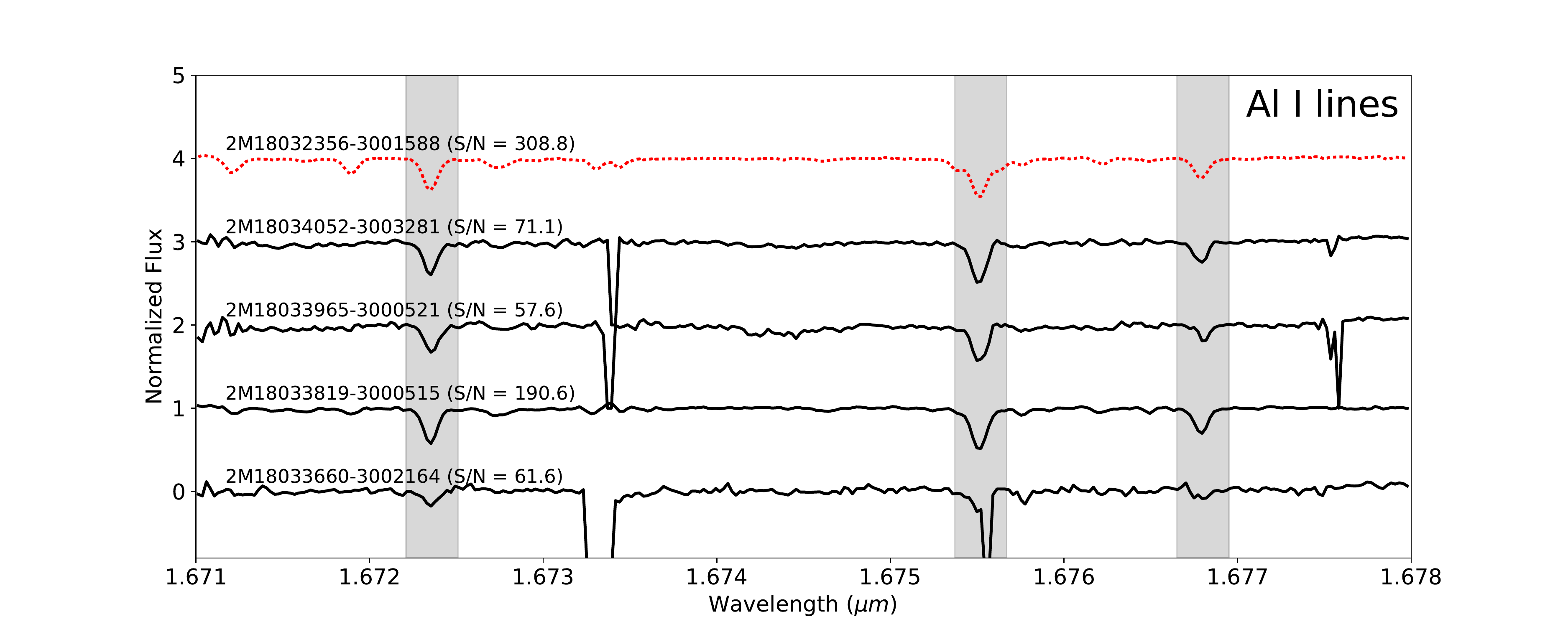}
  		\caption{The APOGEE combined spectra of the analyzed stars in a narrow spectral window, covering the regions (grey shadow) around the CN features (top panel), Mg l lines (middle panel), and Al  I lines (bottom panel) used to estimate N, Mg, and Al abundances. A quick comparison between the stars identified in this work and the APOGEE star (red line) with second-generation abundance patterns identified in a previous paper \citep[][]{Schiavon2017a} indicates that we would be able to detect very large N and Al enhancements. \citep{Alonso2012}}
  		\label{Figure3}
  	\end{center}
  \end{figure*}

  In Figure \ref{Figure3}, we plot several portions of the observed APOGEE spectra, showing examples of the windonws used in our chemical analysis to extract the N, Al, and Mg abundances from the CN lines, Al I, and Mg I spectral features, respectively. The $^{12}$C$^{14}$N and Al I lines are strong for the T$_{\rm eff}$, log \textit{g} and metallicity range of our sample stars, already indicating that they are enhanced in N and Al. The only exception, as expected, is the hottest star in our sample (2M18033660-3002164), which displays much weaker CN, Al I, and Mg I spectral lines in its relatively low-S/N spectrum that makes their abundances more uncertain \citep[in particular for N; see e.g.,][]{Meszaros2015}.
  
  To avoid any spurious results, we rejected the two sodium lines at 1.6373$\mu$m and 1.6388 $\mu$m, as they are very weak in the typical T$_{\rm eff}$, log \textit{g} and metallicity range of our sample, leading to unreliable [Na/Fe] abundances. In addition, lines such as Nd II, Na I, Cr I, Mn I, Ni I, and other chemical species were rejected, as they were found to be weak and heavily blended by other features, which can alter the abundances. 
  
  For each star, the abundances are then derived by means of a line-by-line analysis using the BACCHUS pipeline and MARCS model atmospheres \citep{Gustafsson2008}. The line list adopted in this work is the version linelist.20150714, which was used for the DR14 results \citep{Abolfathi2017}, and it includes both atomic and molecular species. For a more detailed description of these lines, we refer the reader to a forthcoming paper (Holtzman et al. in preparation). 
  
  For each sample star, we need T$_{\rm eff}$ and log \textit{g} as input parameters in BACCHUS. Thus, we decided to use the DR14 ASPCAP uncalibrated effective temperature (T$^{raw}_{\rm eff}$)\footnote{In contrast to \citet{Meszaros2015}, we chose  not  to  estimate the Teff values from any empirical color-temperature relation; this is highly uncertain due to the relatively high NGC 6522 reddening, E(J-K) $\sim$ 0.25 \citep[see e.g.,][]{Schultheis2017}.} that comes from the best ASPCAP global fit to the observed spectra as well as independent surface gravities from PARSEC (Bressan et al. 2012) isochrones (chosen  to  be  12  Gyr). { With fixed T$_{eff}$ and log \textit{g}, the first step consists of determining the metallicity, and $\xi_{t}$ parameter, and the convolution parameter, i.e., the metallicity provided is the average abundance of selected Fe lines, and the $\xi_{t}$ is obtained by minimising the trend of Fe abundances against their reduced equivalent width, while the convolution parameter stands fro the total effect of the instrument resolution, the macroturbulence, and $\nu$ sin \textit{i} on the line broadening} \citep[e.g.,][]{Hawkins2016}. In addition, we have adopted the C, N  and O abundances that satisfy  the  fitting  of  all  molecular lines consistently; i.e., we first derive O abundances from OH, then derive C from CO and N from CN lines and the CNO abundances are derived several times to minimize the OH, CO, and CN dependences \citep[see e.g., ][]{Smith2013, Souto2016}. The  mean  abundances  determined  with  these input atmospheric parameters and the BACCHUS  pipeline  are  listed  in  Table \ref{table2}.
   
  In Table \ref{error1}, we indicate the typical uncertainty of our abundance determinations, i.e., the uncertainty in each of the atmospheric parameters. The final uncerntainty for each element was calculated as the root squared sum of the individual uncertainties due to the errors in each atmospheric parameter under the assumption that these individual uncertainties are independent. The reported uncertainty for each chemical species is: $\sigma_{total}  = \sqrt{\sigma^2_{[X/H], T_{\rm eff}}    + \sigma^2_{[X/H],{\rm log} g} + \sigma^2_{[X/H],\xi_t}  + \sigma^2_{mean}  }$; where $\sigma^2_{mean} $ is calculated using the standard deviation from the different abundances of the different lines for each element, while  $\sigma^2_{[X/H], T_{\rm eff}} $ ,  $\sigma^2_{[X/H],{\rm log} g}$, and $\sigma^2_{[X/H],\xi_t} $ are derived for each chemical specie while varying $T_{eff}$ by $\pm$100 K, log $g$ by $\pm0.3$ dex, and $\epsilon_{t}$ by $\pm0.05$ km s$^{-1}$. These values were chosen as they represent the typical uncertainty in the atmospheric parameters for our sample.

  It is important to note that our line-by-line abundances provides evidence that the new NGC 6522 members reported here are enriched in N and Al, probing the \textit{second-generation} nature of these stars.
  
  \begin{table*}
  	\setlength{\tabcolsep}{0.4mm}  
  	\begin{tiny}
  		\caption{ $G$, $G_{BP}$, $G_{RP}$ and \textit{J}, \textit{H}, \textit{Ks} VVV$+$2MASS magnitudes and kinematics information of the five giant stars analyzed in this work.}
  		\label{table1}
  		\begin{tabular}{cccccccccccccccc}
  			\hline
  			\hline
  			APOGEE ID                               &  $G$ & $G_{BP}$ & $G_{RP}$ &  \textit{J$_{2MASS}$}       &   \textit{H$_{2MASS}$}    &  \textit{Ks$_{2MASS}$} &  \textit{J$_{VVV}$}       &   \textit{H$_{VVV}$}    &  \textit{Ks$_{VVV}$}    & V${\rm helio}$& V$_{\rm scatter}$ & N$_{\rm visits}$ & S/N \\ 
  			&	&	&	&	&	&	&	&	&	& (km s$^{-1}$) & (km s$^{-1}$) & & pixel$^{-1}$\\ 
  			\hline
  			\hline
  			2M18032356$-$3001588     &  12.920  &  14.109    &  11.847      & 10.153$\pm$0.022    &  9.174$\pm$0.025     &  8.936$\pm$0.023    & 10.171$\pm$0.001   &   9.381$\pm$0.001   &   9.157$\pm$0.001    &  $-$13.45$\pm$0.01 &  2.59  & 3 & 308.8\\
  			2M18034052$-$3003281     &  14.789  &  15.679    &  13.781      & 12.109$\pm$0.025    &  11.298$\pm$0.027   &  11.113$\pm$0.027  & 12.291$\pm$0.002   & 11.715$\pm$0.002   & 11.406$\pm$0.002    &  $-$21.97$\pm$0.02 & 0.32  & 7  & 71.1\\
  			2M18033965$-$3000521     &  14.661  &  15.484    &  13.618      & 11.836$\pm$0.034    &  11.104$\pm$0.035   &  10.897$\pm$0.034  & 12.198$\pm$0.002   &   ...                             & 11.329$\pm$0.002    &  $-$19.61$\pm$0.03 & 0.11 & 4 & 57.6 \\
  			2M18033819$-$3000515     &  13.618  &  14.569    &  12.577      & 10.996$\pm$0.028    &  10.106$\pm$0.026   &   9.907$\pm$0.03     & 11.195$\pm$0.001   &  11.201$\pm$0.001   & 10.030$\pm$0.001    &  $-$15.49$\pm$0.01 & 0.05 & 3 & 190.6 \\
  			2M18033660$-$3002164     &  15.322  &  15.818    &  14.144      & 13.006$\pm$0.045    &  11.803                       &  11.574                      & 13.022$\pm$0.004   &  12.443$\pm$0.004   & 12.272$\pm$0.005    & $-$6.61$\pm$0.06 & 0.27 & 7 &  61.6 \\			                                                			                                                
  			\hline
  			\hline
  		\end{tabular} 
  	\end{tiny}
  \end{table*}
  
  \section{Cerium abundances in NGC 6522} 
  \label{neutron}
  
  As we have mentioned above, the two neutron-capture elements Ce and Nd have been
  detected in APOGEE spectra until now \citep[via their Nd II and Ce II H-band
  absorption lines;][]{Hasselquist2016, Cunha2017}, providing an
  unique opportunity to determine the elemental abundances of these elements from
  H-band spectra. Unfortunately, the ten Nd II lines between 15284.5 and 16634.7
  \AA (see Table 3 in Hasselquist et al. 2016) are too weak (and/or heavily
  affected by telluric features) in the APOGEE spectra of our sample stars, being
  not useful for the Nd abundance determination. However, four strong/clean Ce II
  lines (see Table \ref{table3}) are clearly detected in two sample stars (2M18032356-3001588
  and 2M18033819-3000515), which permit us to estimate their Ce abundances. 
  
  The star 2M18032356-3001588 was previously analyzed by \citep{Schiavon2017a} and its Ce abundance ([Ce/Fe]=$+$0.10 dex) has been provided by \citet{Cunha2017}. We measure a BACCHUS-based mean Ce abundance of [Ce/Fe]=$+$0.09$\pm$0.04, which is in excellent agreement with the one reported by \citet{Cunha2017}, while our C, Fe, Al, and Mg abundances agree by $\sim$0.1 dex with the DR12 abundances reported by \citep{Schiavon2017a}; the only exception is N, for which we find a higher N abundance (by 0.25 dex; [N/Fe]=1.29). Thus, 2M18032356-3001588 displays a chemical composition somehow intermediate between the first generation and second-generation stars in the Mg-Al plane as compared to other GCs at similar metallicity (see Figure \ref{Figure5}, left panel).
  
  The star 2M18033819-3000515 shows also a N-enrichment very similar to 2M18032356-3001588. Contrary to 2M18032356-3001588, the star 2M18033819-3000515 displays a mildly enhanced Ce abundance of [Ce/Fe]=+0.23$\pm$0.03, which is accompanied by a higher Al content (and lower Mg) that is consistent with a \textit{second-generation} nature.
  
  For 2M18033965$-$3000521, 2M18034052$-$3003281 and 2M18033660$-$3002164, the Ce II absortion lines are heavily affected by telluric features and too weak to be derived, and were not well reproduced by the synthesis. Thus, we do not provide the [Ce/Fe] abundance ratios for these stars.

  \section{Discussion}
  \label{results}
  
\begin{figure*}
    	\begin{center}
    		\includegraphics[width=200mm]{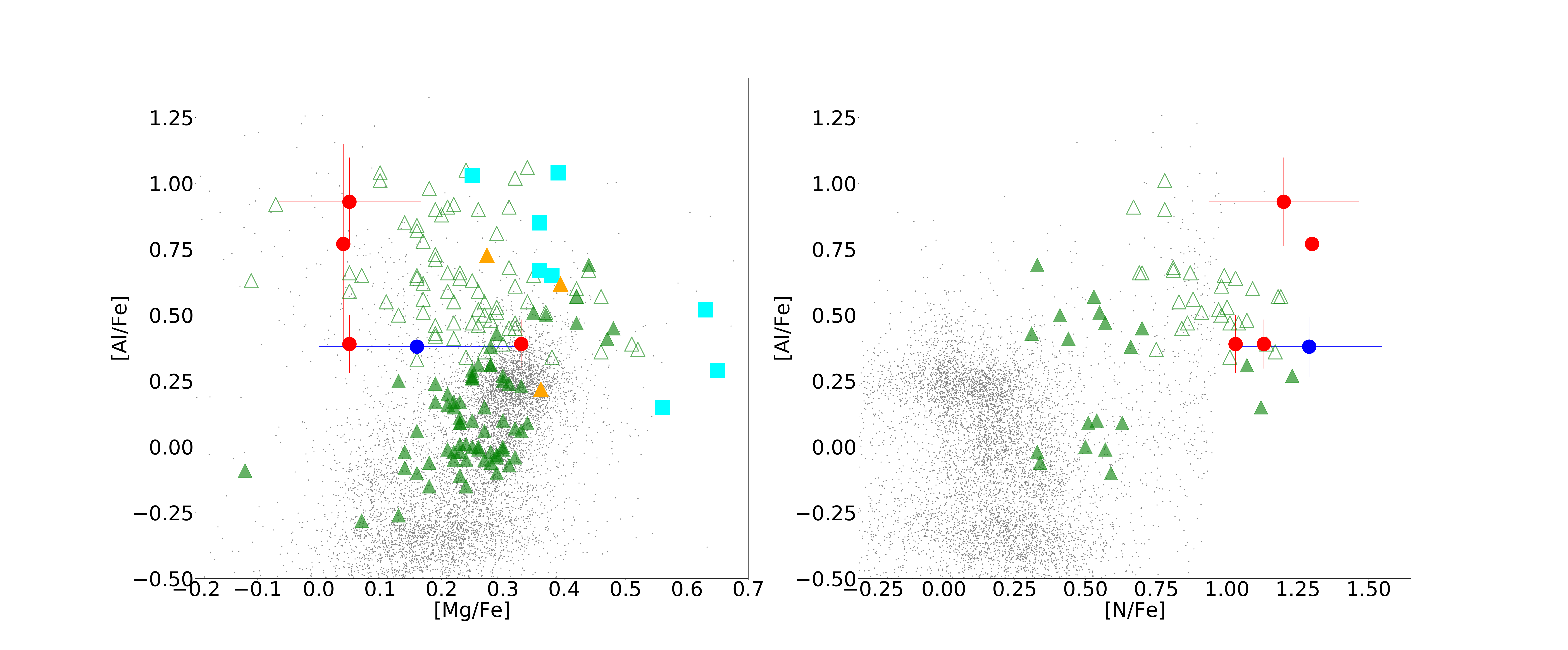}
    		\caption{ The behaviour of the average [Al/Fe], [Mg/Fe], and [N/Fe] abundance ratios of our synthesis analysis (red and blue filled symbols) compared with DR14 abundances from field stars (grey dots), and overplotted with APOGEE DR14 determinations for the first- (green filled triangles) and second-populations (green empty triangles) in GCs, M5, M71, and M107 \citep{Meszaros2015}. Orange triangles and cyan squares are very likely members of NGC 6522 from the Gaia-ESO survey \citet{Recio-Blanco2017} and \citet{Ness2014}, respectively.}
    		\label{Figure5}
    	\end{center}
    \end{figure*}

  Two of the stars analyzed in the present sample (2M18033819$-$3000515 and 2M18034052$-$3003281) show high Al abundances ([Al/Fe]$> +0.77$), potentially associated with a second stellar generation. This is also corroborated by the high N ([N/Fe]$> +1.0$), indicating a clear correlation between these two elements. This is in good agreement with the self-enrichment scenario where the origin of the SG chemical pattern is attributed to the pollution with gas reprocessed by proton-capture nucleosynthesis \citep[see][]{Meszaros2015}. { The other three stars in the sample (2M18032356$-$3001588, 2M18033965$-$3000521, and 2M18033660-3002164) exhibit lower Al enhancement ($\sim +0.4$ dex) with respect to the solar-scaled Al-abundance, while they are clearly highly N enhanced ([N/Fe]$> +1.0$), and occupy the locus dominated by second-generation globular cluster stars at similar metallicity, and separated relatively cleanly in the [N/Fe]--[Fe/H] plane; see Figure \ref{Figure5}. }
 
 { 
 We caution on the accuracy of [Al/Fe] for 2M18033660-3002164, whose Al I line in $\lambda^{\rm air} = 16710$ \AA{} is weaker; while it has a high N abundance, we warn that these lines are nor reliable.  For $^{12}$C$^{16}$O and $^{16}$OH the lines are weak and heavily blended by telluric features. At this time, we cannot guarantee the quality of the [C/Fe] and [O/Fe] abundances for 2M18033660-3002164, this is not the case for $^{12}$C$^{14}$N and therefore the [N/Fe] abundance ratio have been derived by fixing A(C) and A(O) using the reported [O/Fe] and [C/Fe] from \citet{Chiappini2011};  star B$-$107 (2M18033660-3002164) in that work.}
    
  Any conclusion given on the basis of the Mg abundances is less trivial. The small size of the APOGEE sample discussed here limits the possibility of clearly identifying stars of the first generation (FG). This affects, in particular, any conclusion on the presence or not of Mg depletion in this cluster based solely on APOGEE data. In clusters of similar metallicities, the Mg variation between FG and SG members is generally smaller \citep[$<= +0.2$ dex, see, e.g.][]{Meszaros2015} than what is observed in Al and N. More caution must be taken when considering that such Mg variation is comparable with the abundance uncertainties. Nevertheless, the APOGEE data suggest the presence of a Mg-Al anticorrelation. This Mg-Al anticorrelation has also been observed by \citep{Ness2014}, where abundances for a larger sample of stars (8) have been measured. From Figure \ref{Figure5}, however, the Mg measurements from Ness et al. seem to be systematically higher than the present APOGEE sample and the \citet{Recio-Blanco2017} sample. This is also confirmed by \citet{Barbuy2014}, where they found [Mg/Fe] systematically higher ($\sim $+0.2 dex) for 4 stars in common with the \citet{Ness2014} sample. A larger stellar sample, analyzed in a homogeneous fashion, with more accurate abundances, is needed to further confirm the presence of a possible Mg spread between FG and SG stars. This is also what is needed in order to directly compare these observations with any GC formation/evolution scenario so far proposed to explain the origin of the multiple populations \citep[see, e.g.,][for a general review]{Bastian2018}. Most the stars in our final dataset lie in a group with super-solar [N/Fe] and [Al/Fe], and clearly extend beyond of the typical chemical abundances observed in Milky Way field stars.

  The [O/Fe] abundance ratios, listed in Table \ref{table2}, are generally higher compared with APOGEE-DR14/ASPCAP results, by $\sim+$0.15 dex, showing that [O/Fe] abundance ratios are particularly sensitive to log \textit{g}. As the abundances of C and O affect CN lines \citep[see][]{Schiavon2017a}, it can seen in Table \ref{table2} that the variations in [O/Fe] does not affect significantly the [N/Fe] abundance ratios in our sample, which turn out to be nitrogen rich, with remarkably stronger CN lines with non-enhanced carbon abundances ([C/Fe] $\lesssim$ +0.15). In other words, these stars exhibit clear N enhancements, even when [O/Fe] is slightly sensitive to log \textit{g}. 
  
\begin{figure}
    	\begin{center}
    		\includegraphics[width=100mm]{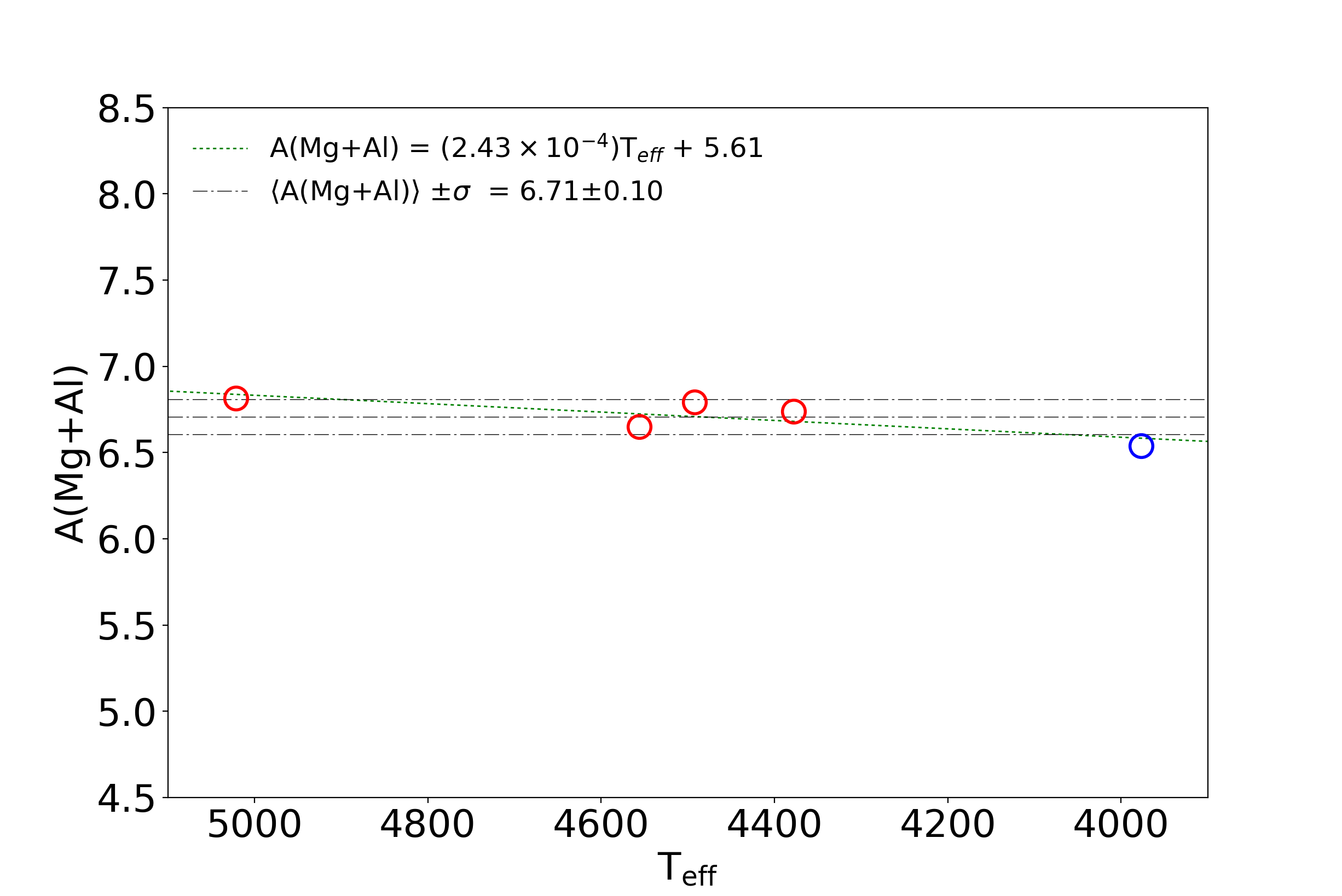}
    		\caption{Combined abundance of A(Mg+Al) as a function of effective temperature (T$_{\rm eff}$). The symbols have the same meaning as those in Figure \ref{Figure1}.}
    		\label{Figure9}
    	\end{center}
    \end{figure}

  As mentioned above, the newly identified stellar members of NGC 6522 display enhancements in [Al/Fe], suggesting that NGC 6522 exhibits large scatter in its Al abundance ratios. Combining our results with the abundances analysis from \citet{Ness2014} and \citet{Recio-Blanco2017}, we infer Al variations to be $\Delta$[Al/Fe]$\sim$ 1 dex. Such Al enhancements provide an indication that multiple populations with distinctive chemistry are present in NGC 6522, and that the MgAl cycles have been activated. Figure \ref{Figure5} clearly shows the Mg-Al anti-correlation in our sample, and the [Mg/Fe] abundances show a much smaller variation ($\Delta$[Mg/Fe]$\lesssim$+0.2 dex) in our MARCS/BACCHUS determinations. However, the combined datasets show that Mg exhibits significantly larger scatter than any implicit systematic error. This comparison allows us to confirm the conversion of Mg into Al during the MgAl cycles, which is present in NGC 6522. { The summed abundance A(Mg+Al) is expected to be constant as a function of T$_{\rm eff}$ when material is completely processed through the MgAl cycle, and that is what our results show in Figure \ref{Figure9}. } This finding is a clear confirmation of the results reported in our previous work \citep[see][]{Schiavon2017a, Recio-Blanco2017}. 
  
  Concerning silicon, we found over-abundances of [Si/Fe] ratios, on the order of $\sim +$0.3, which is similar to APOGEE-DR14/ASPCAP values, with a reasonably small scatter, within our measurement errors. So far, our  abundance values fall into acceptable ranges with the literature on abundance studies in globular cluster stars \citep[e.g., see][]{Carretta2012, Meszaros2015, Recio-Blanco2017}. We find that the Si-Al correlation is also weak in our data. This could be interpreted as evidence for Si leaking from the Mg-Al cycle \citep[for discussion and references, see, e.g., ][]{Tang2017}, i.e., one would expect the Si enhancement to be correlated with Al in metal-poor globular clusters, \textit{where the AGB stars burn slightly hotter or in high-mass clusters, where the chemical enrichment is more efficient} \citep[see][]{Carretta2009a, Meszaros2015}.
  
  \textit{Radial velocity variation:} The stars 2M18033819$-$3000515, 2M18033965$-$3000521, 2M18034052$-$3003281, and 2M18033660$-$3002164 were visited 3, 4, 7, and 7 times, respectively, by the APOGEE survey. This allow us to identify any significant variation in their radial velocities, in order to add empirical constraints to the origin of the observed N and Al over-abundances. Thus, given that the typical variation in radial velocity measured for these stars is of the order of V$_{scatter} <$ 0.4 km s$^{-1}$ \citep[see][]{Nidever2015}, this rules out the binary mass-transfer hypothesis \cite[see][]{Schiavon2017b} as possible polluters.  
  
  It is important to note that the derived [Ce/Fe] ratios are compatible with previous works that found low s-process abundances, as \citep[e.g.,][]{Ness2014}. Our results reinforce the hypothesis that the \textit{s-}process rich material in NGC6522 could have been formed due to the pollution of the pristine gas by a former population of massive AGB stars \citep[e.g.,][]{Ventura2016a, Flavia2017,Fishlock2014} and, on the other side, they do not support an scenario in which the spinstars are the main polluters.
  
  Finally, \citet{Fernandez-Trincado2017a, Fernandez-Trincado2017b, Fernandez-Trincado2019c} have recently discovered a new N- and Al-rich ([N/Fe] and [Al/Fe] ratios around $\sim$+1.0 dex) population of stars on very eccentric orbits (\textit{e}$>$0.65) in the Milky Way field (towards the bulge, the disk, and the halo), passing through the inner regions of the Milky Way bulge.  Whether globular clusters at similar metallicities are able to kick out stars with similar chemical behavior, as seen in the innermost regions of NGC 6522, we would expect that a few field stars with similar chemistry patterns \citep[][for instance]{Schiavon2017b, Fernandez-Trincado2017a, Fernandez-Trincado2019a, Fernandez-Trincado2019b, Fernandez-Trincado2019c, Fernandez-Trincado2019d} could have been ejected from these bulge cluster enviroments with a relative velocity greater that the escape velocity of the GCs, particularly being ejected from some scenarios involving binary systems or black hole interactions \citep[see, e.g.,][]{Hut1983, Heggie1996, Pichardo2012, Fernandez-Trincado2013, Fernandez-Trincado2015a, Fernandez-Trincado2015b, Fernandez-Trincado2016a}, or due to simple tidal forces \citep{Kupper2012, Lane2012}. In turn, these could be capable of exceeding the escape velocity at the radius of the bulge ($\sim 650$ km s$^{-1}$), this means that we would expect a few of the N-rich stars with enhanced Al abundances ([Al/Fe]$\gtrsim$+0.6) not to be part of the Milky Way bulge and would therefore describe eccentric orbits, as recently found. More accurate distances and proper motions are needed to confirm this hypothesis.

  \section{Concluding remarks} 
  \label{section5}
  
  We have used an independent pipeline called BACCHUS \citep[see][]{Masseron2016, Hawkins2016}, an updated line list and careful line selection to explore the chemical abundance patterns of five potential members of the globular cluster NGC 6522. 
  
  The distinctive chemical patterns characterising multiple populations, specially enrichment in nitrogen and aluminium, simultaneous with low carbon-abundance ratios ([C/Fe]$<$ +0.15) have been measured in our sample, thus confirming the presence of multiple populations in NGC 6522.
  
  The main results of our chemical abundance analysis from high-resolution APOGEE spectra in NGC 6522 potential members can be summarised as follows:
  
  \begin{itemize}
  	
  	\item We report the identification of three new potential stellar members (2M18033819-3000515, 2M18033965-3000521 and 2M18034052-3003281) of NGC 6522 in the Apache Point Observatory Galactic Evolution Experiment (APOGEE) survey \citep{Majewski2017}. The spectra analyzed in this work have a signal-to-noise ratios larger than 50, exhibiting very similar line strengths (namely CN bands, Al I and Mg I lines) to that of 2M18032356-3001588 \citep[see][]{Schiavon2017b},
  	making also them ideal for line-by-line spectrum synthesis calculations of selected clean features. These spectral properties suggest that this group of stars share a common formation history, and spatial relationship on the sky, and are therefore gravitationally bound to NGC 6522.
  	
  	\item We have measured significant N and Al over-abundances, with carbon depletion in NGC 6522 members, suggesting that the distinctive chemical patterns characterising multiple stellar populations is present within NGC 6522, it reinforces the recent claims in the literature \citep{Schiavon2017a, Recio-Blanco2017, Kerber2018}. 
  	
  	\item Lastly, we do not find any enhancement in heavy elements measured from APOGEE spectrum (Ce II). We have measured only mildly enhanced [Ce/Fe]$<$0.25 abundance ratios, in agreement with recent optical studies, which contradict previous observational evidence for the chemical signatures of rapidly rotating Population III stars ("spinstars") in NGC 6522. Such low \textit{s}-process abundances could still be consistent with other intra-cluster medium polluters such as massive AGB stars.
  	
  \end{itemize}

\begin{table*}
	\setlength{\tabcolsep}{5.0mm}  
	\begin{tiny}
		\caption{ Mean abundance as derived from BACCHUS for elements which have more than 1 line.}
		\label{table2}
		\begin{tabular}{cccccc}
			\hline
			\hline
			APOGEE ID                                &  2M18032356-        &   2M18033819-     &  2M18033965-           & 2M18034052-         & 2M18033660-            \\ 
			&  3001588                 &   3000515             &  3000521                     & 3003281                  & 3002164                    \\ 
			\hline
			\hline
			T$_{\rm eff}$  (K)                         & 3977.2                      & 4378.1                    & 4555.8                       & 4492.3                   &  5021.1                       \\
			log \textit{g}    (dex)                       & 0.50                          & 1.09                         &  1.35                         & 1.26                       &   1.99                          \\
			$\xi_t$  (km s$^{-1}$)    &  2.46 & 2.67  &  2.43 &  1.92 & 1.37 \\
			$ {\rm [Fe/H] }  $                     & $-$1.20     & $-$0.97       &  $-$0.99     &  $-$0.97  &  $-$1.08       \\
			$ {\rm [C/Fe]  } $                     & $-$0.48     & $-$0.33       &  $-$0.29     &  $-$0.24   &              ...                   \\
			$ {\rm [N/Fe]  } $                     &   1.29     &  1.30        &   1.03      &   1.20   &  1.13       \\
			$ {\rm [O/Fe]  } $                     &   0.39     &  0.30        &   0.29      &  0.35     &            ...                    \\
			
			$ {\rm [Al/Fe] }  $                     &  0.38      &    0.77      &  0.39     &     0.93   &  0.39    \\
			$ {\rm [Mg/Fe]} $                     &  0.16       &    0.04      & 0.05  &     0.05      &  0.33    \\
			$ {\rm [Si/Fe] }  $                     &  0.28       &    0.17      &  0.18   &     0.37     &  0.51    \\
			$ {\rm [Ce/Fe] }  $                    &  0.09      &    0.23       &  ...    &     ...      &   ...     \\
			\hline 
			\hline
		\end{tabular} 
		\tablefoot{The Solar reference abundances are from \citet{Asplund2005}, except for Ce, which is taken from \citet{Grevesse2015}.}
	\end{tiny}
\end{table*}

\begin{table}
	\begin{tiny}
		\begin{center}
			\setlength{\tabcolsep}{1.0mm}  
			\caption{ Typical uncertainty of the abundance determinations from our present measurements.}
			\begin{tabular}{clccccc}
				\hline
				\hline
				APOGEE$-$ID                    &   $X$ &   $\sigma_{[X/H],T_{\rm eff}}$   &    $\sigma_{[X/H],logg}$   &   $\sigma_{[X/H],\xi_{t}}$  &   $\sigma_{mean}$    &   $\sigma_{total}$  \\
				\hline
				\hline
				2M18032356$-$3001588 &   Fe       &   0.055   &   0.033     &  0.041     &   0.050  &   0.091    \\
				2M18032356$-$3001588  &   C       &   0.038   &   0.132    &    0.081    &   0.030  &   0.162   \\
				2M18032356$-$3001588  &   N       &   0.172    &  0.135     &   0.122    &   0.060  &   0.257    \\
				2M18032356$-$3001588  &   O       &   0.141   &   0.075    &    0.076   &   0.070  &    0.190   \\
				2M18032356$-$3001588  &   Mg     &   0.092   &   0.097    &    0.071   &   0.050  &   0.159    \\
				2M18032356$-$3001588  &   Al      &    0.075    &  0.053     &   0.067    &  0.010  &  0.114     \\
				2M18032356$-$3001588  &   Si       &    0.022   &   0.017    &    0.021    &  0.100  & 0.106      \\
				2M18032356$-$3001588  &   Ce     &    0.044   &   0.072    &    0.039   &   0.040  &    0.101   \\
				\hline
				2M18034052$-$3003281 &   Fe       &  0.054    &   0.035    &  0.024     &  0.080   &   0.105    \\
				2M18034052$-$3003281 &   C        &  0.084    &   0.121    &  0.040     &  0.020   &   0.154    \\
				2M18034052$-$3003281  &   N       &  0.190    &   0.151    &  0.039     &  0.100   &   0.265     \\
				2M18034052$-$3003281  &   O       &  0.146    &   0.049    &  0.017     &  0.020   &   0.156   \\
				2M18034052$-$3003281  &   Mg     &  0.060    &   0.066    &  0.044     &  0.060   &   0.116     \\
				2M18034052$-$3003281  &   Al      &  0.118     &   0.098    &  0.031     &  0.060   &  0.168     \\
				2M18034052$-$3003281  &   Si       &  0.033     &  0.025    &  0.014     &  0.090   &  0.100     \\
				2M18034052$-$3003281  &   Ce     &    ...          &    ....       &    ...         &  ...         &   ...    \\				
				\hline
				2M18033965$-$3000521 &   Fe       &   0.032   &    0.041   &  0.008     &  0.040   &    0.066   \\
				2M18033965$-$3000521  &   C       &   0.071   &     0.092   &  0.116    &  0.040   &    0.169   \\
				2M18033965$-$3000521  &   N       &   0.121   &     0.071  &   0.149    &  0.050   &    0.211  \\
				2M18033965$-$3000521  &   O       &   0.121   &     0.053  &   0.027    &  0.020   &    0.136   \\
				2M18033965$-$3000521  &   Mg     &   0.068   &     0.037  &   0.019    &  0.050   &    0.094   \\
				2M18033965$-$3000521  &   Al      &   0.099    &     0.041  &   0.021    &  0.02   &       0.111\\
				2M18033965$-$3000521  &   Si       &   0.032   &     0.067  &   0.024    &   0.06  &      0.098 \\
				2M18033965$-$3000521  &   Ce     &    ...         &     ....       &    ...      &   ...     &       \\		
				\hline
				2M18033819$-$3000515 &   Fe       &  0.061    &  0.092     &   0.011    &  0.050   & 0.122      \\
				2M18033819$-$3000515  &   C       &  0.020    &   0.133    &   0.017    &  0.040   &  0.141     \\
				2M18033819$-$3000515  &   N       &  0.148    &   0.234    &   0.010    &  0.050   &  0.282     \\
				2M18033819$-$3000515  &   O       &  0.149    &   0.025    &   0.003    &  0.070   &   0.166    \\
				2M18033819$-$3000515  &   Mg     &  0.128    &   0.201    &   0.034    &  0.080   &   0.254    \\
				2M18033819$-$3000515 &   Al       &  0.244     &   0.277    &   0.037    &  0.070   &   0.378   \\
				2M18033819$-$3000515  &   Si       &  0.163    &    0.214    &   0.046    &  0.080   &   0.284    \\
				2M18033819$-$3000515  &   Ce     &   0.045    &    0.143    &   0.004    &  0.030   &    0.153   \\
				\hline
				2M18033660$-$3002164 &   Fe       &  0.049     &  0.018     &  0.037     &  0.100   &    0.119   \\
				2M18033660$-$3002164  &   C       &  ...    & ...      &   ...    &  ...       &     ...  \\
				2M18033660$-$3002164  &   N       &  0.247    &    0.103   &   0.059    &  0.130   &   0.303    \\
				2M18033660$-$3002164 &   O       &    ...  &    ...   &  ...     &   ...       &  ...     \\
				2M18033660$-$3002164  &   Mg     &  0.071    &    0.077   &    0.044   &  0.150   &   0.188    \\
				2M18033660$-$3002164  &   Al      &  0.065     &    0.033   &   0.057    &  ...      &   0.093    \\
				2M18033660$-$3002164  &   Si       &  0.043    &     0.018  &    0.025   &  0.290   &  0.295     \\
				2M18033660$-$3002164  &   Ce     &   ...     &  ...     &   ...    &  ...       &   ...    \\								
				\hline		
				\hline
			\end{tabular}  \label{error1}\\
		\end{center}
	\end{tiny}
\end{table}

\clearpage
\newpage

\begin{acknowledgements} 
	
We would like to thank the referee for insightful comments that helped to improve this work. J.G.F-T, P.L-P, and J.A-G were supported by MINEDUC-UA project, code ANT 1855. J.G.F-T  also acknowledges financial support from the FONDECYT No. 3180210 and the ChETEC COST Action (CA16117), supported by COST (European Cooperation in Science and Technology). D.G. gratefully acknowledges support from the Chilean Centro de Excelencia en Astrof\'isica y Tecnolog\'ias Afines (CATA) BASAL grant AFB-170002. D.G. also acknowledges financial support from the Direcci\'on de Investigaci\'on y Desarrollo de la Universidad de La Serena through the Programa de Incentivo a la Investigaci\'on de Acad\'emicos (PIA-DIDULS). S.V gratefully acknowledges the support provided by Fondecyt reg. n. 1170518. Szabolcs M{\'e}sz{\'a}ros has been supported by the Premium Postdoctoral Research Program of the Hungarian Academy of Sciences, and by the Hungarian NKFI Grants K-119517 of the Hungarian National Research, Development and Innovation Office. R.E.M. acknowledges project fondecyt 1190621. D.M. and J.A-G. are supported also by FONDECYT No. 1170121 and 11150916 respectively, and by the Ministry of Economy, Development, and Tourism’s Millennium Science Initiative through grant IC120009, awarded to the Millennium Institute of Astrophysics (MAS). DAGH, OZ, FDA, and TM acknowledge support from the State Research Agency (AEI) of the Spanish Ministry of Science, Innovation and Universities (MCIU) and the European Regional Development Fund (FEDER) under grant AYA2017-88254-P. T.C.B. acknowledges partial support from grant PHY 14-30152: Physics Frontier Center/JINA Center for the Evolution of the Elements (JINA-CEE), awarded by the US National Science Foundation.\newline

\texttt{BACCHUS} have been executed on computers from the Institute of Astronomy and Planetary Sciences at Universidad de Atacama.\newline

This work has made use of results from the European Space Agency (ESA) space mission {\it Gaia}, the data from which were processed by the {\it Gaia Data Processing and Analysis Consortium} (DPAC).  Funding for the DPAC has been provided by national institutions, in particular the institutions participating in the {\it Gaia} Multilateral Agreement. The {\it Gaia} mission website is \url{http: //www.cosmos.esa.int/gaia}. Funding for the Sloan Digital Sky Survey IV has been provided by the Alfred P. Sloan Foundation, the U.S. Department of Energy Office of Science, and the Participating Institutions. SDSS- IV acknowledges support and resources from the Center for High-Performance Computing at the University of Utah. The SDSS web site is www.sdss.org. SDSS-IV is managed by the Astrophysical Research Consortium for the Participating Institutions of the SDSS Collaboration including the Brazilian Participation Group, the Carnegie Institution for Science, Carnegie Mellon University, the Chilean Participation Group, the French Participation Group, Harvard-Smithsonian Center for Astrophysics, Instituto de Astrof\`{i}sica de Canarias, The Johns Hopkins University, Kavli Institute for the Physics and Mathematics of the Universe (IPMU) / University of Tokyo, Lawrence Berkeley National Laboratory, Leibniz Institut f\"{u}r Astrophysik Potsdam (AIP), Max-Planck-Institut f\"{u}r Astronomie (MPIA Heidelberg), Max-Planck-Institut f\"{u}r Astrophysik (MPA Garching), Max-Planck-Institut f\"{u}r Extraterrestrische Physik (MPE), National Astronomical Observatory of China, New Mexico State University, New York University, University of  Dame, Observat\'{o}rio Nacional / MCTI, The Ohio State University, Pennsylvania State University, Shanghai Astronomical Observatory, United Kingdom Participation Group, Universidad Nacional Aut\'{o}noma de M\'{e}xico, University of Arizona, University of Colorado Boulder, University of Oxford, University of Portsmouth, University of Utah, University of Virginia, University of Washington, University of Wisconsin, Vanderbilt University, and Yale University.

\end{acknowledgements}


\begin{thebibliography}{79}
	\expandafter\ifx\csname natexlab\endcsname\relax\def\natexlab#1{#1}\fi
	
	\bibitem[{{Abolfathi} {et~al.}(2018){Abolfathi}, {Aguado}, {Aguilar}, {Allende
			Prieto}, {Almeida}, {Ananna}, {Anders}, {Anderson}, {Andrews}, {Anguiano}, \&
		et~al.}]{Abolfathi2017}
	{Abolfathi}, B., {Aguado}, D.~S., {Aguilar}, G., {et~al.} 2018, \apjs, 235, 42
	
	\bibitem[{{Alam} {et~al.}(2015){Alam}, {Albareti}, {Allende Prieto}, {Anders},
		{Anderson}, {Anderton}, {Andrews}, {Armengaud}, {Aubourg}, {Bailey}, \&
		et~al.}]{Alam2015}
	{Alam}, S., {Albareti}, F.~D., {Allende Prieto}, C., {et~al.} 2015, \apjs, 219,
	12
	
	\bibitem[{{Albareti} {et~al.}(2017){Albareti}, {Allende Prieto}, {Almeida},
		{Anders}, {Anderson}, {Andrews}, {Arag{\'o}n-Salamanca},
		{Argudo-Fern{\'a}ndez}, {Armengaud}, {Aubourg}, \& et~al.}]{Albareti2017}
	{Albareti}, F.~D., {Allende Prieto}, C., {Almeida}, A., {et~al.} 2017, \apjs,
	233, 25
	
	\bibitem[{{Alonso-Garc{\'{\i}}a} {et~al.}(2012){Alonso-Garc{\'{\i}}a}, {Mateo},
		{Sen}, {Banerjee}, {Catelan}, {Minniti}, \& {von Braun}}]{Alonso2012}
	{Alonso-Garc{\'{\i}}a}, J., {Mateo}, M., {Sen}, B., {et~al.} 2012, \aj, 143, 70
	
	\bibitem[{{Asplund} {et~al.}(2005){Asplund}, {Grevesse}, \&
		{Sauval}}]{Asplund2005}
	{Asplund}, M., {Grevesse}, N., \& {Sauval}, A.~J. 2005, in Astronomical Society
	of the Pacific Conference Series, Vol. 336, Cosmic Abundances as Records of
	Stellar Evolution and Nucleosynthesis, ed. T.~G. {Barnes}, III \& F.~N.
	{Bash}, 25
	
	\bibitem[{{Barbuy} {et~al.}(2014){Barbuy}, {Chiappini}, {Cantelli}, {Depagne},
		{Pignatari}, {Hirschi}, {Cescutti}, {Ortolani}, {Hill}, {Zoccali}, {Minniti},
		{Trevisan}, {Bica}, \& {G{\'o}mez}}]{Barbuy2014}
	{Barbuy}, B., {Chiappini}, C., {Cantelli}, E., {et~al.} 2014, \aap, 570, A76
	
	\bibitem[{{Barbuy} {et~al.}(2009){Barbuy}, {Zoccali}, {Ortolani}, {Hill},
		{Minniti}, {Bica}, {Renzini}, \& {G{\'o}mez}}]{Barbuy2009}
	{Barbuy}, B., {Zoccali}, M., {Ortolani}, S., {et~al.} 2009, \aap, 507, 405
	
	\bibitem[{{Bastian} \& {Lardo}(2018)}]{Bastian2018}
	{Bastian}, N. \& {Lardo}, C. 2018, \araa, 56, 83
	
	\bibitem[{{Carretta}(2016)}]{Carretta2016}
	{Carretta}, E. 2016, ArXiv e-prints [\eprint[arXiv]{1611.04728}]
	
	\bibitem[{{Carretta} {et~al.}(2009{\natexlab{a}}){Carretta}, {Bragaglia},
		{Gratton}, \& {Lucatello}}]{Carretta2009a}
	{Carretta}, E., {Bragaglia}, A., {Gratton}, R., \& {Lucatello}, S.
	2009{\natexlab{a}}, \aap, 505, 139
	
	\bibitem[{{Carretta} {et~al.}(2010){Carretta}, {Bragaglia}, {Gratton},
		{Lucatello}, {Bellazzini}, \& {D'Orazi}}]{Carretta2010}
	{Carretta}, E., {Bragaglia}, A., {Gratton}, R., {et~al.} 2010, \apjl, 712, L21
	
	\bibitem[{{Carretta} {et~al.}(2009{\natexlab{b}}){Carretta}, {Bragaglia},
		{Gratton}, {Lucatello}, {Catanzaro}, {Leone}, {Bellazzini}, {Claudi},
		{D'Orazi}, {Momany}, {Ortolani}, {Pancino}, {Piotto}, {Recio-Blanco}, \&
		{Sabbi}}]{Carretta2009b}
	{Carretta}, E., {Bragaglia}, A., {Gratton}, R.~G., {et~al.} 2009{\natexlab{b}},
	\aap, 505, 117
	
	\bibitem[{{Carretta} {et~al.}(2012){Carretta}, {Bragaglia}, {Gratton},
		{Lucatello}, \& {D'Orazi}}]{Carretta2012}
	{Carretta}, E., {Bragaglia}, A., {Gratton}, R.~G., {Lucatello}, S., \&
	{D'Orazi}, V. 2012, \apjl, 750, L14
	
	\bibitem[{{Carretta} {et~al.}(2007){Carretta}, {Bragaglia}, {Gratton},
		{Momany}, {Recio-Blanco}, {Cassisi}, {Fran{\c c}ois}, {James}, {Lucatello},
		\& {Moehler}}]{Carretta2007}
	{Carretta}, E., {Bragaglia}, A., {Gratton}, R.~G., {et~al.} 2007, \aap, 464,
	967
	
	\bibitem[{{Chiappini} {et~al.}(2011){Chiappini}, {Frischknecht}, {Meynet},
		{Hirschi}, {Barbuy}, {Pignatari}, {Decressin}, \& {Maeder}}]{Chiappini2011}
	{Chiappini}, C., {Frischknecht}, U., {Meynet}, G., {et~al.} 2011, \nat, 472,
	454
	
	\bibitem[{{Cohen} {et~al.}(2017){Cohen}, {Moni Bidin}, {Mauro}, {Bonatto}, \&
		{Geisler}}]{Cohen2017a}
	{Cohen}, R.~E., {Moni Bidin}, C., {Mauro}, F., {Bonatto}, C., \& {Geisler}, D.
	2017, \mnras, 464, 1874
	
	\bibitem[{{Cunha} {et~al.}(2017){Cunha}, {Smith}, {Hasselquist}, {Souto},
		{Shetrone}, {Allende Prieto}, {Bizyaev}, {Frinchaboy},
		{Garc{\'{\i}}a-Hern{\'a}ndez}, {Holtzman}, {Johnson}, {J{\H o}nsson},
		{Majewski}, {M{\'e}sz{\'a}ros}, {Nidever}, {Pinsonneault}, {Schiavon},
		{Sobeck}, {Skrutskie}, {Zamora}, {Zasowski}, \&
		{Fern{\'a}ndez-Trincado}}]{Cunha2017}
	{Cunha}, K., {Smith}, V.~V., {Hasselquist}, S., {et~al.} 2017, \apj, 844, 145
	
	\bibitem[{{Dell'Agli} {et~al.}(2018){Dell'Agli}, {Garc{\'{\i}}a-Hern{\'a}ndez},
		{Ventura}, {M{\'e}sz{\'a}ros}, {Masseron}, {Fern{\'a}ndez-Trincado}, {Tang},
		{Shetrone}, {Zamora}, \& {Lucatello}}]{Flavia2017}
	{Dell'Agli}, F., {Garc{\'{\i}}a-Hern{\'a}ndez}, D.~A., {Ventura}, P., {et~al.}
	2018, \mnras, 475, 3098
	
	\bibitem[{{Eisenstein} {et~al.}(2011){Eisenstein}, {Weinberg}, {Agol},
		{Aihara}, {Allende Prieto}, {Anderson}, {Arns}, {Aubourg}, {Bailey},
		{Balbinot}, \& et~al.}]{Eisenstein2011}
	{Eisenstein}, D.~J., {Weinberg}, D.~H., {Agol}, E., {et~al.} 2011, \aj, 142, 72
	
	\bibitem[{{Fern{\'a}ndez-Trincado}
		{et~al.}(2019{\natexlab{a}}){Fern{\'a}ndez-Trincado}, {Beers}, {Placco},
		{Martell}, {Moreno}, {Tang}, {Alves-Brito}, {Villanova}, {Ortigoza-Urdaneta},
		{Minniti}, {P{\'e}rez-Villegas}, {Reyl{\'e}}, \&
		{Robin}}]{Fernandez-Trincado2019d}
	{Fern{\'a}ndez-Trincado}, J.~G., {Beers}, T.~C., {Placco}, V.~M., {et~al.}
	2019{\natexlab{a}}, arXiv e-prints [\eprint[arXiv]{1904.05884}]
	
	\bibitem[{{Fern{\'a}ndez-Trincado}
		{et~al.}(2019{\natexlab{b}}){Fern{\'a}ndez-Trincado}, {Beers}, {Tang},
		{Moreno}, {P{\'e}rez-Villegas}, \&
		{Ortigoza-Urdaneta}}]{Fernandez-Trincado2019b}
	{Fern{\'a}ndez-Trincado}, J.~G., {Beers}, T.~C., {Tang}, B., {et~al.}
	2019{\natexlab{b}}, arXiv e-prints [\eprint[arXiv]{1904.05369}]
	
	\bibitem[{{Fern{\'a}ndez-Trincado}
		{et~al.}(2017{\natexlab{a}}){Fern{\'a}ndez-Trincado}, {Geisler}, {Moreno},
		{Zamora}, {Robin}, \& {Villanova}}]{Fernandez-Trincado2017b}
	{Fern{\'a}ndez-Trincado}, J.~G., {Geisler}, D., {Moreno}, E., {et~al.}
	2017{\natexlab{a}}, in SF2A-2017: Proceedings of the Annual meeting of the
	French Society of Astronomy and Astrophysics, ed. C.~{Reyl{\'e}}, P.~{Di
		Matteo}, F.~{Herpin}, E.~{Lagadec}, A.~{Lan{\c c}on}, Z.~{Meliani}, \&
	F.~{Royer}, 199--202
	
	\bibitem[{{Fern{\'a}ndez-Trincado}
		{et~al.}(2019{\natexlab{c}}){Fern{\'a}ndez-Trincado}, {Mennickent},
		{Cabezas}, {Zamora}, {Martell}, {Beers}, {Placco}, {Nataf},
		{M{\'e}sz{\'a}ros}, {Minniti}, {Schleicher}, {Tang}, {P{\'e}rez-Villegas},
		{Robin}, \& {Reyl{\'e}}}]{Fernandez-Trincado2019a}
	{Fern{\'a}ndez-Trincado}, J.~G., {Mennickent}, R., {Cabezas}, M., {et~al.}
	2019{\natexlab{c}}, arXiv e-prints [\eprint[arXiv]{1902.10635}]
	
	\bibitem[{{Fern{\'a}ndez-Trincado}
		{et~al.}(2019{\natexlab{d}}){Fern{\'a}ndez-Trincado}, {Ortigoza-Urdaneta},
		{Moreno}, {P{\'e}rez-Villegas}, \& {Soto}}]{Fernandez-Trincado2019c}
	{Fern{\'a}ndez-Trincado}, J.~G., {Ortigoza-Urdaneta}, M., {Moreno}, E.,
	{P{\'e}rez-Villegas}, A., \& {Soto}, M. 2019{\natexlab{d}}, arXiv e-prints
	[\eprint[arXiv]{1904.05370}]
	
	\bibitem[{{Fern{\'a}ndez-Trincado}
		{et~al.}(2016{\natexlab{a}}){Fern{\'a}ndez-Trincado}, {Robin}, {Moreno},
		{Schiavon}, {Garc{\'{\i}}a P{\'e}rez}, {Vieira}, {Cunha}, {Zamora}, {Sneden},
		{Souto}, {Carrera}, {Johnson}, {Shetrone}, {Zasowski},
		{Garc{\'{\i}}a-Hern{\'a}ndez}, {Majewski}, {Reyl{\'e}}, {Blanco-Cuaresma},
		{Martinez-Medina}, {P{\'e}rez-Villegas}, {Valenzuela}, {Pichardo}, {Meza},
		{M{\'e}sz{\'a}ros}, {Sobeck}, {Geisler}, {Anders}, {Schultheis}, {Tang},
		{Roman-Lopes}, {Mennickent}, {Pan}, {Nitschelm}, \&
		{Allard}}]{Fernandez-Trincado2016b}
	{Fern{\'a}ndez-Trincado}, J.~G., {Robin}, A.~C., {Moreno}, E., {et~al.}
	2016{\natexlab{a}}, \apj, 833, 132
	
	\bibitem[{{Fern{\'a}ndez-Trincado}
		{et~al.}(2016{\natexlab{b}}){Fern{\'a}ndez-Trincado}, {Robin}, {Reyl{\'e}},
		{Vieira}, {Palmer}, {Moreno}, {Valenzuela}, \&
		{Pichardo}}]{Fernandez-Trincado2016a}
	{Fern{\'a}ndez-Trincado}, J.~G., {Robin}, A.~C., {Reyl{\'e}}, C., {et~al.}
	2016{\natexlab{b}}, \mnras, 461, 1404
	
	\bibitem[{{Fern{\'a}ndez-Trincado}
		{et~al.}(2015{\natexlab{a}}){Fern{\'a}ndez-Trincado}, {Robin}, {Vieira},
		{Moreno}, {Bienaym{\'e}}, {Reyl{\'e}}, {Valenzuela}, {Pichardo},
		{Robles-Valdez}, \& {Martins}}]{Fernandez-Trincado2015b}
	{Fern{\'a}ndez-Trincado}, J.~G., {Robin}, A.~C., {Vieira}, K., {et~al.}
	2015{\natexlab{a}}, \aap, 583, A76
	
	\bibitem[{{Fern{\'a}ndez Trincado} {et~al.}(2013){Fern{\'a}ndez Trincado},
		{Vivas}, {Mateu}, \& {Zinn}}]{Fernandez-Trincado2013}
	{Fern{\'a}ndez Trincado}, J.~G., {Vivas}, A.~K., {Mateu}, C.~E., \& {Zinn}, R.
	2013, \memsai, 84, 265
	
	\bibitem[{{Fern{\'a}ndez-Trincado}
		{et~al.}(2015{\natexlab{b}}){Fern{\'a}ndez-Trincado}, {Vivas}, {Mateu},
		{Zinn}, {Robin}, {Valenzuela}, {Moreno}, \&
		{Pichardo}}]{Fernandez-Trincado2015a}
	{Fern{\'a}ndez-Trincado}, J.~G., {Vivas}, A.~K., {Mateu}, C.~E., {et~al.}
	2015{\natexlab{b}}, \aap, 574, A15
	
	\bibitem[{{Fern{\'a}ndez-Trincado}
		{et~al.}(2017{\natexlab{b}}){Fern{\'a}ndez-Trincado}, {Zamora},
		{Garc{\'{\i}}a-Hern{\'a}ndez}, {Souto}, {Dell'Agli}, {Schiavon}, {Geisler},
		{Tang}, {Villanova}, {Hasselquist}, {Mennickent}, {Cunha}, {Shetrone},
		{Allende Prieto}, {Vieira}, {Zasowski}, {Sobeck}, {Hayes}, {Majewski},
		{Placco}, {Beers}, {Schleicher}, {Robin}, {M{\'e}sz{\'a}ros}, {Masseron},
		{Garc{\'{\i}}a P{\'e}rez}, {Anders}, {Meza}, {Alves-Brito}, {Carrera},
		{Minniti}, {Lane}, {Fern{\'a}ndez-Alvar}, {Moreno}, {Pichardo},
		{P{\'e}rez-Villegas}, {Schultheis}, {Roman-Lopes}, {Fuentes}, {Nitschelm},
		{Harding}, {Bizyaev}, {Pan}, {Oravetz}, {Simmons}, {Ivans},
		{Blanco-Cuaresma}, {Hern{\'a}ndez}, {Alonso-Garc{\'{\i}}a}, {Valenzuela}, \&
		{Chanam{\'e}}}]{Fernandez-Trincado2017a}
	{Fern{\'a}ndez-Trincado}, J.~G., {Zamora}, O., {Garc{\'{\i}}a-Hern{\'a}ndez},
	D.~A., {et~al.} 2017{\natexlab{b}}, \apjl, 846, L2
	
	\bibitem[{{Fishlock} {et~al.}(2014){Fishlock}, {Karakas}, {Lugaro}, \&
		{Yong}}]{Fishlock2014}
	{Fishlock}, C.~K., {Karakas}, A.~I., {Lugaro}, M., \& {Yong}, D. 2014, \apj,
	797, 44
	
	\bibitem[{{Gaia Collaboration} {et~al.}(2018){Gaia Collaboration}, {Brown},
		{Vallenari}, {Prusti}, {de Bruijne}, {Babusiaux}, {Bailer-Jones}, {Biermann},
		{Evans}, {Eyer}, \& et~al.}]{gaiadr2}
	{Gaia Collaboration}, {Brown}, A.~G.~A., {Vallenari}, A., {et~al.} 2018, AAP,
	616, A1
	
	\bibitem[{{Garc{\'{\i}}a-Hern{\'a}ndez}
		{et~al.}(2015){Garc{\'{\i}}a-Hern{\'a}ndez}, {M{\'e}sz{\'a}ros}, {Monelli},
		{Cassisi}, {Stetson}, {Zamora}, {Shetrone}, \&
		{Lucatello}}]{Garcia-Hernandez2015}
	{Garc{\'{\i}}a-Hern{\'a}ndez}, D.~A., {M{\'e}sz{\'a}ros}, S., {Monelli}, M.,
	{et~al.} 2015, \apjl, 815, L4
	
	\bibitem[{{Garc{\'{\i}}a P{\'e}rez} {et~al.}(2016){Garc{\'{\i}}a P{\'e}rez},
		{Allende Prieto}, {Holtzman}, {Shetrone}, {M{\'e}sz{\'a}ros}, {Bizyaev},
		{Carrera}, {Cunha}, {Garc{\'{\i}}a-Hern{\'a}ndez}, {Johnson}, {Majewski},
		{Nidever}, {Schiavon}, {Shane}, {Smith}, {Sobeck}, {Troup}, {Zamora},
		{Weinberg}, {Bovy}, {Eisenstein}, {Feuillet}, {Frinchaboy}, {Hayden},
		{Hearty}, {Nguyen}, {O'Connell}, {Pinsonneault}, {Wilson}, \&
		{Zasowski}}]{GarciaPerez2016a}
	{Garc{\'{\i}}a P{\'e}rez}, A.~E., {Allende Prieto}, C., {Holtzman}, J.~A.,
	{et~al.} 2016, \aj, 151, 144
	
	\bibitem[{{Gnedin} \& {Ostriker}(1997)}]{Gnedin1997}
	{Gnedin}, O.~Y. \& {Ostriker}, J.~P. 1997, \apj, 474, 223
	
	\bibitem[{{Gratton} {et~al.}(2004){Gratton}, {Sneden}, \&
		{Carretta}}]{Gratton2004}
	{Gratton}, R., {Sneden}, C., \& {Carretta}, E. 2004, \araa, 42, 385
	
	\bibitem[{{Gratton} {et~al.}(2012){Gratton}, {Carretta}, \&
		{Bragaglia}}]{Gratton2012}
	{Gratton}, R.~G., {Carretta}, E., \& {Bragaglia}, A. 2012, \aapr, 20, 50
	
	\bibitem[{{Gratton} {et~al.}(2007){Gratton}, {Lucatello}, {Bragaglia},
		{Carretta}, {Cassisi}, {Momany}, {Pancino}, {Valenti}, {Caloi}, {Claudi},
		{D'Antona}, {Desidera}, {Fran{\c c}ois}, {James}, {Moehler}, {Ortolani},
		{Pasquini}, {Piotto}, \& {Recio-Blanco}}]{Gratton2007}
	{Gratton}, R.~G., {Lucatello}, S., {Bragaglia}, A., {et~al.} 2007, \aap, 464,
	953
	
	\bibitem[{{Grevesse} {et~al.}(2015){Grevesse}, {Scott}, {Asplund}, \&
		{Sauval}}]{Grevesse2015}
	{Grevesse}, N., {Scott}, P., {Asplund}, M., \& {Sauval}, A.~J. 2015, \aap, 573,
	A27
	
	\bibitem[{{Gunn} {et~al.}(2006){Gunn}, {Siegmund}, {Mannery}, {Owen}, {Hull},
		{Leger}, {Carey}, {Knapp}, {York}, {Boroski}, {Kent}, {Lupton}, {Rockosi},
		{Evans}, {Waddell}, {Anderson}, {Annis}, {Barentine}, {Bartoszek}, {Bastian},
		{Bracker}, {Brewington}, {Briegel}, {Brinkmann}, {Brown}, {Carr},
		{Czarapata}, {Drennan}, {Dombeck}, {Federwitz}, {Gillespie}, {Gonzales},
		{Hansen}, {Harvanek}, {Hayes}, {Jordan}, {Kinney}, {Klaene}, {Kleinman},
		{Kron}, {Kresinski}, {Lee}, {Limmongkol}, {Lindenmeyer}, {Long}, {Loomis},
		{McGehee}, {Mantsch}, {Neilsen}, {Neswold}, {Newman}, {Nitta}, {Peoples},
		{Pier}, {Prieto}, {Prosapio}, {Rivetta}, {Schneider}, {Snedden}, \&
		{Wang}}]{Gunn2006}
	{Gunn}, J.~E., {Siegmund}, W.~A., {Mannery}, E.~J., {et~al.} 2006, \aj, 131,
	2332
	
	\bibitem[{{Gustafsson} {et~al.}(2008){Gustafsson}, {Edvardsson}, {Eriksson},
		{J{\o}rgensen}, {Nordlund}, \& {Plez}}]{Gustafsson2008}
	{Gustafsson}, B., {Edvardsson}, B., {Eriksson}, K., {et~al.} 2008, \aap, 486,
	951
	
	\bibitem[{{Harris}(1996)}]{Harris1996}
	{Harris}, W.~E. 1996, \aj, 112, 1487
	
	\bibitem[{{Hasselquist} {et~al.}(2016){Hasselquist}, {Shetrone}, {Cunha},
		{Smith}, {Holtzman}, {Lawler}, {Allende Prieto}, {Beers}, {Chojnowski},
		{Fern{\'a}ndez-Trincado}, {Garc{\'{\i}}a-Hern{\'a}ndez}, {Hearty},
		{Majewski}, {Pereira}, {Placco}, {Villanova}, \& {Zamora}}]{Hasselquist2016}
	{Hasselquist}, S., {Shetrone}, M., {Cunha}, K., {et~al.} 2016, \apj, 833, 81
	
	\bibitem[{{Hawkins} {et~al.}(2016){Hawkins}, {Masseron}, {Jofr{\'e}},
		{Gilmore}, {Elsworth}, \& {Hekker}}]{Hawkins2016}
	{Hawkins}, K., {Masseron}, T., {Jofr{\'e}}, P., {et~al.} 2016, \aap, 594, A43
	
	\bibitem[{{Heggie} {et~al.}(1996){Heggie}, {Hut}, \& {McMillan}}]{Heggie1996}
	{Heggie}, D.~C., {Hut}, P., \& {McMillan}, S.~L.~W. 1996, \apj, 467, 359
	
	\bibitem[{{Hut}(1983)}]{Hut1983}
	{Hut}, P. 1983, \apjl, 272, L29
	
	\bibitem[{{Kerber} {et~al.}(2018){Kerber}, {Nardiello}, {Ortolani}, {Barbuy},
		{Bica}, {Cassisi}, {Libralato}, \& {Vieira}}]{Kerber2018}
	{Kerber}, L.~O., {Nardiello}, D., {Ortolani}, S., {et~al.} 2018, \apj, 853, 15
	
	\bibitem[{{K{\"u}pper} {et~al.}(2012){K{\"u}pper}, {Lane}, \&
		{Heggie}}]{Kupper2012}
	{K{\"u}pper}, A.~H.~W., {Lane}, R.~R., \& {Heggie}, D.~C. 2012, \mnras, 420,
	2700
	
	\bibitem[{{Lane} {et~al.}(2012){Lane}, {K{\"u}pper}, \& {Heggie}}]{Lane2012}
	{Lane}, R.~R., {K{\"u}pper}, A.~H.~W., \& {Heggie}, D.~C. 2012, \mnras, 423,
	2845
	
	\bibitem[{{Lee} {et~al.}(2019){Lee}, {Kim}, {Johnson}, {Chung}, {Jang}, {Lim},
		\& {Kang}}]{Lee2019}
	{Lee}, Y.-W., {Kim}, J.~J., {Johnson}, C.~I., {et~al.} 2019, \apj, 878, L2
	
	\bibitem[{{Majewski} {et~al.}(2017){Majewski}, {Schiavon}, {Frinchaboy},
		{Allende Prieto}, {Barkhouser}, {Bizyaev}, {Blank}, {Brunner}, {Burton},
		{Carrera}, {Chojnowski}, {Cunha}, {Epstein}, {Fitzgerald}, {Garc{\'{\i}}a
			P{\'e}rez}, {Hearty}, {Henderson}, {Holtzman}, {Johnson}, {Lam}, {Lawler},
		{Maseman}, {M{\'e}sz{\'a}ros}, {Nelson}, {Nguyen}, {Nidever}, {Pinsonneault},
		{Shetrone}, {Smee}, {Smith}, {Stolberg}, {Skrutskie}, {Walker}, {Wilson},
		{Zasowski}, {Anders}, {Basu}, {Beland}, {Blanton}, {Bovy}, {Brownstein},
		{Carlberg}, {Chaplin}, {Chiappini}, {Eisenstein}, {Elsworth}, {Feuillet},
		{Fleming}, {Galbraith-Frew}, {Garc{\'{\i}}a}, {Garc{\'{\i}}a-Hern{\'a}ndez},
		{Gillespie}, {Girardi}, {Gunn}, {Hasselquist}, {Hayden}, {Hekker}, {Ivans},
		{Kinemuchi}, {Klaene}, {Mahadevan}, {Mathur}, {Mosser}, {Muna}, {Munn},
		{Nichol}, {O'Connell}, {Parejko}, {Robin}, {Rocha-Pinto}, {Schultheis},
		{Serenelli}, {Shane}, {Silva Aguirre}, {Sobeck}, {Thompson}, {Troup},
		{Weinberg}, \& {Zamora}}]{Majewski2017}
	{Majewski}, S.~R., {Schiavon}, R.~P., {Frinchaboy}, P.~M., {et~al.} 2017, \aj,
	154, 94
	
	\bibitem[{{Martell} \& {Grebel}(2010)}]{Martell2010}
	{Martell}, S.~L. \& {Grebel}, E.~K. 2010, \aap, 519, A14
	
	\bibitem[{{Martell} {et~al.}(2016){Martell}, {Shetrone}, {Lucatello},
		{Schiavon}, {M{\'e}sz{\'a}ros}, {Allende Prieto},
		{Garc{\'{\i}}a-Hern{\'a}ndez}, {Beers}, \& {Nidever}}]{Martell2016}
	{Martell}, S.~L., {Shetrone}, M.~D., {Lucatello}, S., {et~al.} 2016, \apj, 825,
	146
	
	\bibitem[{{Martell} {et~al.}(2011){Martell}, {Smolinski}, {Beers}, \&
		{Grebel}}]{Martell2011}
	{Martell}, S.~L., {Smolinski}, J.~P., {Beers}, T.~C., \& {Grebel}, E.~K. 2011,
	\aap, 534, A136
	
	\bibitem[{{Masseron} {et~al.}(2016){Masseron}, {Merle}, \&
		{Hawkins}}]{Masseron2016}
	{Masseron}, T., {Merle}, T., \& {Hawkins}, K. 2016, {BACCHUS: Brussels
		Automatic Code for Characterizing High accUracy Spectra}, Astrophysics Source
	Code Library
	
	\bibitem[{{M{\'e}sz{\'a}ros} {et~al.}(2018){M{\'e}sz{\'a}ros},
		{Garc{\'{\i}}a-Hern{\'a}ndez}, {Cassisi}, {Monelli}, {Szigeti}, {Dell'Agli},
		{Derekas}, {Masseron}, {Shetrone}, {Stetson}, \& {Zamora}}]{Meszaros2017}
	{M{\'e}sz{\'a}ros}, S., {Garc{\'{\i}}a-Hern{\'a}ndez}, D.~A., {Cassisi}, S.,
	{et~al.} 2018, \mnras, 475, 1633
	
	\bibitem[{{M{\'e}sz{\'a}ros} {et~al.}(2015){M{\'e}sz{\'a}ros}, {Martell},
		{Shetrone}, {Lucatello}, {Troup}, {Bovy}, {Cunha},
		{Garc{\'{\i}}a-Hern{\'a}ndez}, {Overbeek}, {Allende Prieto}, {Beers},
		{Frinchaboy}, {Garc{\'{\i}}a P{\'e}rez}, {Hearty}, {Holtzman}, {Majewski},
		{Nidever}, {Schiavon}, {Schneider}, {Sobeck}, {Smith}, {Zamora}, \&
		{Zasowski}}]{Meszaros2015}
	{M{\'e}sz{\'a}ros}, S., {Martell}, S.~L., {Shetrone}, M., {et~al.} 2015, \aj,
	149, 153
	
	\bibitem[{{Minniti} {et~al.}(2010){Minniti}, {Lucas}, {Emerson}, {Saito},
		{Hempel}, {Pietrukowicz}, {Ahumada}, {Alonso}, {Alonso-Garcia}, {Arias},
		{Bandyopadhyay}, {Barb{\'a}}, {Barbuy}, {Bedin}, {Bica}, {Borissova},
		{Bronfman}, {Carraro}, {Catelan}, {Clari{\'a}}, {Cross}, {de Grijs},
		{D{\'e}k{\'a}ny}, {Drew}, {Fari{\~n}a}, {Feinstein}, {Fern{\'a}ndez
			Laj{\'u}s}, {Gamen}, {Geisler}, {Gieren}, {Goldman}, {Gonzalez}, {Gunthardt},
		{Gurovich}, {Hambly}, {Irwin}, {Ivanov}, {Jord{\'a}n}, {Kerins}, {Kinemuchi},
		{Kurtev}, {L{\'o}pez-Corredoira}, {Maccarone}, {Masetti}, {Merlo},
		{Messineo}, {Mirabel}, {Monaco}, {Morelli}, {Padilla}, {Palma}, {Parisi},
		{Pignata}, {Rejkuba}, {Roman-Lopes}, {Sale}, {Schreiber}, {Schr{\"o}der},
		{Smith}, {}, {Soto}, {Tamura}, {Tappert}, {Thompson}, {Toledo}, {Zoccali}, \&
		{Pietrzynski}}]{Minniti2010}
	{Minniti}, D., {Lucas}, P.~W., {Emerson}, J.~P., {et~al.} 2010, \na, 15, 433
	
	\bibitem[{{Mu{\~n}oz} {et~al.}(2017){Mu{\~n}oz}, {Villanova}, {Geisler},
		{Saviane}, {Dias}, {Cohen}, \& {Mauro}}]{Cesar2017}
	{Mu{\~n}oz}, C., {Villanova}, S., {Geisler}, D., {et~al.} 2017, \aap, 605, A12
	
	\bibitem[{{Ness} {et~al.}(2014){Ness}, {Asplund}, \& {Casey}}]{Ness2014}
	{Ness}, M., {Asplund}, M., \& {Casey}, A.~R. 2014, \mnras, 445, 2994
	
	\bibitem[{{Nidever} {et~al.}(2015){Nidever}, {Holtzman}, {Allende Prieto},
		{Beland}, {Bender}, {Bizyaev}, {Burton}, {Desphande}, {Fleming},
		{Garc{\'{\i}}a P{\'e}rez}, {Hearty}, {Majewski}, {M{\'e}sz{\'a}ros}, {Muna},
		{Nguyen}, {Schiavon}, {Shetrone}, {Skrutskie}, {Sobeck}, \&
		{Wilson}}]{Nidever2015}
	{Nidever}, D.~L., {Holtzman}, J.~A., {Allende Prieto}, C., {et~al.} 2015, \aj,
	150, 173
	
	\bibitem[{{Pancino} {et~al.}(2017){Pancino}, {Romano}, {Tang}, {Tautvai{\v
				s}ien{\.e}}, {Casey}, {Gruyters}, {Geisler}, {San Roman}, {Randich},
		{Alfaro}, {Bragaglia}, {Flaccomio}, {Korn}, {Recio-Blanco}, {Smiljanic},
		{Carraro}, {Bayo}, {Costado}, {Damiani}, {Jofr{\'e}}, {Lardo}, {de Laverny},
		{Monaco}, {Morbidelli}, {Sbordone}, {Sousa}, \& {Villanova}}]{Pancino2017}
	{Pancino}, E., {Romano}, D., {Tang}, B., {et~al.} 2017, \aap, 601, A112
	
	\bibitem[{{Pichardo} {et~al.}(2012){Pichardo}, {Moreno}, {Allen}, {Bedin},
		{Bellini}, \& {Pasquini}}]{Pichardo2012}
	{Pichardo}, B., {Moreno}, E., {Allen}, C., {et~al.} 2012, \aj, 143, 73
	
	\bibitem[{{Pignatari} {et~al.}(2008){Pignatari}, {Gallino}, {Meynet},
		{Hirschi}, {Herwig}, \& {Wiescher}}]{Pignatari2008}
	{Pignatari}, M., {Gallino}, R., {Meynet}, G., {et~al.} 2008, \apjl, 687, L95
	
	\bibitem[{{Recio-Blanco} {et~al.}(2017){Recio-Blanco}, {Rojas-Arriagada}, {de
			Laverny}, {Mikolaitis}, {Hill}, {Zoccali}, {Fern{\'a}ndez-Trincado}, {Robin},
		{Babusiaux}, {Gilmore}, {Randich}, {Alfaro}, {Allende Prieto}, {Bragaglia},
		{Carraro}, {Jofr{\'e}}, {Lardo}, {Monaco}, {Morbidelli}, \&
		{Zaggia}}]{Recio-Blanco2017}
	{Recio-Blanco}, A., {Rojas-Arriagada}, A., {de Laverny}, P., {et~al.} 2017,
	\aap, 602, L14
	
	\bibitem[{{Schiavon} {et~al.}(2017{\natexlab{a}}){Schiavon}, {Johnson},
		{Frinchaboy}, {Zasowski}, {M{\'e}sz{\'a}ros}, {Garc{\'{\i}}a-Hern{\'a}ndez},
		{Cohen}, {Tang}, {Villanova}, {Geisler}, {Beers}, {Fern{\'a}ndez-Trincado},
		{Garc{\'{\i}}a P{\'e}rez}, {Lucatello}, {Majewski}, {Martell}, {O'Connell},
		{Prieto}, {Bizyaev}, {Carrera}, {Lane}, {Malanushenko}, {Malanushenko},
		{Mu{\~n}oz}, {Nitschelm}, {Oravetz}, {Pan}, {Roman-Lopes}, {Schultheis}, \&
		{Simmons}}]{Schiavon2017a}
	{Schiavon}, R.~P., {Johnson}, J.~A., {Frinchaboy}, P.~M., {et~al.}
	2017{\natexlab{a}}, \mnras, 466, 1010
	
	\bibitem[{{Schiavon} {et~al.}(2017{\natexlab{b}}){Schiavon}, {Zamora},
		{Carrera}, {Lucatello}, {Robin}, {Ness}, {Martell}, {Smith},
		{Garc{\'{\i}}a-Hern{\'a}ndez}, {Manchado}, {Sch{\"o}nrich}, {Bastian},
		{Chiappini}, {Shetrone}, {Mackereth}, {Williams}, {M{\'e}sz{\'a}ros},
		{Allende Prieto}, {Anders}, {Bizyaev}, {Beers}, {Chojnowski}, {Cunha},
		{Epstein}, {Frinchaboy}, {Garc{\'{\i}}a P{\'e}rez}, {Hearty}, {Holtzman},
		{Johnson}, {Kinemuchi}, {Majewski}, {Muna}, {Nidever}, {Nguyen}, {O'Connell},
		{Oravetz}, {Pan}, {Pinsonneault}, {Schneider}, {Schultheis}, {Simmons},
		{Skrutskie}, {Sobeck}, {Wilson}, \& {Zasowski}}]{Schiavon2017b}
	{Schiavon}, R.~P., {Zamora}, O., {Carrera}, R., {et~al.} 2017{\natexlab{b}},
	\mnras, 465, 501
	
	\bibitem[{{Schultheis} {et~al.}(2017){Schultheis}, {Rojas-Arriagada},
		{Garc{\'{\i}}a P{\'e}rez}, {J{\"o}nsson}, {Hayden}, {Nandakumar}, {Cunha},
		{Allende Prieto}, {Holtzman}, {Beers}, {Bizyaev}, {Brinkmann}, {Carrera},
		{Cohen}, {Geisler}, {Hearty}, {Fernandez-Tricado}, {Maraston}, {Minnitti},
		{Nitschelm}, {Roman-Lopes}, {Schneider}, {Tang}, {Villanova}, {Zasowski}, \&
		{Majewski}}]{Schultheis2017}
	{Schultheis}, M., {Rojas-Arriagada}, A., {Garc{\'{\i}}a P{\'e}rez}, A.~E.,
	{et~al.} 2017, \aap, 600, A14
	
	\bibitem[{{Skrutskie} {et~al.}(2006){Skrutskie}, {Cutri}, {Stiening},
		{Weinberg}, {Schneider}, {Carpenter}, {Beichman}, {Capps}, {Chester},
		{Elias}, {Huchra}, {Liebert}, {Lonsdale}, {Monet}, {Price}, {Seitzer},
		{Jarrett}, {Kirkpatrick}, {Gizis}, {Howard}, {Evans}, {Fowler}, {Fullmer},
		{Hurt}, {Light}, {Kopan}, {Marsh}, {McCallon}, {Tam}, {Van Dyk}, \&
		{Wheelock}}]{Skrutskie2006}
	{Skrutskie}, M.~F., {Cutri}, R.~M., {Stiening}, R., {et~al.} 2006, \aj, 131,
	1163
	
	\bibitem[{{Smith} {et~al.}(2013){Smith}, {Cunha}, {Shetrone}, {Meszaros},
		{Allende Prieto}, {Bizyaev}, {Garc{\`i}a P{\`e}rez}, {Majewski}, {Schiavon},
		{Holtzman}, \& {Johnson}}]{Smith2013}
	{Smith}, V.~V., {Cunha}, K., {Shetrone}, M.~D., {et~al.} 2013, \apj, 765, 16
	
	\bibitem[{{Souto} {et~al.}(2016){Souto}, {Cunha}, {Smith}, {Allende Prieto},
		{Pinsonneault}, {Zamora}, {Garc{\'{\i}}a-Hern{\'a}ndez}, {M{\'e}sz{\'a}ros},
		{Bovy}, {Garc{\'{\i}}a P{\'e}rez}, {Anders}, {Bizyaev}, {Carrera},
		{Frinchaboy}, {Holtzman}, {Ivans}, {Majewski}, {Shetrone}, {Sobeck}, {Pan},
		{Tang}, {Villanova}, \& {Geisler}}]{Souto2016}
	{Souto}, D., {Cunha}, K., {Smith}, V., {et~al.} 2016, \apj, 830, 35
	
	\bibitem[{{Tang} {et~al.}(2017){Tang}, {Cohen}, {Geisler}, {Schiavon},
		{Majewski}, {Villanova}, {Carrera}, {Zamora}, {Garcia-Hernandez}, {Shetrone},
		{Frinchaboy}, {Meza}, {Fern{\'a}ndez-Trincado}, {Mu{\~n}oz}, {Lin}, {Lane},
		{Nitschelm}, {Pan}, {Bizyaev}, {Oravetz}, \& {Simmons}}]{Tang2017}
	{Tang}, B., {Cohen}, R.~E., {Geisler}, D., {et~al.} 2017, \mnras, 465, 19
	
     \bibitem[{{Tang} {et~al.}(2018){Tang}, {Fern{\'a}ndez-Trincado}, {Geisler}, {Zamora},
			{Majewski}, {Villanova}, {Carrera}, {Zamora}, {Garcia-Hernandez}, {Shetrone},
			{Frinchaboy}, {Meza}, {Fern{\'a}ndez-Trincado}, {Mu{\~n}oz}, {Lin}, {Lane},
			{Nitschelm}, {Pan}, {Bizyaev}, {Oravetz}, \& {Simmons}}]{Tang2017}
		{Tang}, B., {Cohen}, R.~E., {Geisler}, D., {et~al.} 2018, \apj, 855, 38
		
	\bibitem[{{Tang} {et~al.}(2019){Tang}, {Liu}, {Fern{\'a}ndez-Trincado},
		{Geisler}, {Shi}, {Zamora}, {Worthey}, \& {Moreno}}]{Tang2019}
	{Tang}, B., {Liu}, C., {Fern{\'a}ndez-Trincado}, J.~G., {et~al.} 2019, \apj,
	871, 58
	
	\bibitem[{{Vasiliev}(2019)}]{Vasiliev18}
	{Vasiliev}, E. 2019, \mnras, 484, 2832
	
	\bibitem[{{Ventura} {et~al.}(2016){Ventura}, {Garc{\'{\i}}a-Hern{\'a}ndez},
		{Dell'Agli}, {D'Antona}, {M{\'e}sz{\'a}ros}, {Lucatello}, {Di Criscienzo},
		{Shetrone}, {Tailo}, {Tang}, \& {Zamora}}]{Ventura2016a}
	{Ventura}, P., {Garc{\'{\i}}a-Hern{\'a}ndez}, D.~A., {Dell'Agli}, F., {et~al.}
	2016, \apjl, 831, L17
	
	\bibitem[{{Wilson} {et~al.}(2012){Wilson}, {Hearty}, {Skrutskie}, {Majewski},
		{Schiavon}, {Eisenstein}, {Gunn}, {Holtzman}, {Nidever}, {Gillespie},
		{Weinberg}, {Blank}, {Henderson}, {Smee}, {Barkhouser}, {Harding}, {Hope},
		{Fitzgerald}, {Stolberg}, {Arns}, {Nelson}, {Brunner}, {Burton}, {Walker},
		{Lam}, {Maseman}, {Barr}, {Leger}, {Carey}, {MacDonald}, {Ebelke}, {Beland},
		{Horne}, {Young}, {Rieke}, {Rieke}, {O'Brien}, {Crane}, {Carr}, {Harrison},
		{Stoll}, {Vernieri}, {Shetrone}, {Allende-Prieto}, {Johnson}, {Frinchaboy},
		{Zasowski}, {Garcia Perez}, {Bizyaev}, {Cunha}, {Smith}, {Meszaros}, {Zhao},
		{Hayden}, {Chojnowski}, {Andrews}, {Loomis}, {Owen}, {Klaene}, {Brinkmann},
		{Stauffer}, {Long}, {Jordan}, {Holder}, {Cope}, {Naugle}, {Pfaffenberger},
		{Schlegel}, {Blanton}, {Muna}, {Weaver}, {Snedden}, {Pan}, {Brewington},
		{Malanushenko}, {Malanushenko}, {Simmons}, {Oravetz}, {Mahadevan}, \&
		{Halverson}}]{Wilson2012}
	{Wilson}, J.~C., {Hearty}, F., {Skrutskie}, M.~F., {et~al.} 2012, in \procspie,
	Vol. 8446, Ground-based and Airborne Instrumentation for Astronomy IV, 84460H
	
	\bibitem[{{Zasowski} {et~al.}(2017){Zasowski}, {Cohen}, {Chojnowski},
		{Santana}, {Oelkers}, {Andrews}, {Beaton}, {Bender}, {Bird}, {Bovy},
		{Carlberg}, {Covey}, {Cunha}, {Dell'Agli}, {Fleming}, {Frinchaboy},
		{Garc{\'{\i}}a-Hern{\'a}ndez}, {Harding}, {Holtzman}, {Johnson}, {Kollmeier},
		{Majewski}, {M{\'e}sz{\'a}ros}, {Munn}, {Mu{\~n}oz}, {Ness}, {Nidever},
		{Poleski}, {Rom{\'a}n-Z{\'u}{\~n}iga}, {Shetrone}, {Simon}, {Smith},
		{Sobeck}, {Stringfellow}, {Szigeti{\'a}ros}, {Tayar}, \&
		{Troup}}]{Zasowski2017}
	{Zasowski}, G., {Cohen}, R.~E., {Chojnowski}, S.~D., {et~al.} 2017, \aj, 154,
	198
	
	\bibitem[{{Zasowski} {et~al.}(2013){Zasowski}, {Johnson}, {Frinchaboy},
		{Majewski}, {Nidever}, {Rocha Pinto}, {Girardi}, {Andrews}, {Chojnowski},
		{Cudworth}, {Jackson}, {Munn}, {Skrutskie}, {Beaton}, {Blake}, {Covey},
		{Deshpande}, {Epstein}, {Fabbian}, {Fleming}, {Garcia Hernandez}, {Herrero},
		{Mahadevan}, {M{\'e}sz{\'a}ros}, {Schultheis}, {Sellgren}, {Terrien}, {van
			Saders}, {Allende Prieto}, {Bizyaev}, {Burton}, {Cunha}, {da Costa},
		{Hasselquist}, {Hearty}, {Holtzman}, {Garc{\'{\i}}a P{\'e}rez}, {Maia},
		{O'Connell}, {O'Donnell}, {Pinsonneault}, {Santiago}, {Schiavon}, {Shetrone},
		{Smith}, \& {Wilson}}]{Zasowski2013}
	{Zasowski}, G., {Johnson}, J.~A., {Frinchaboy}, P.~M., {et~al.} 2013, \aj, 146,
	81
	
\end{thebibliography}

\begin{appendix}
	
	\section{Line-by-line abundance determination}
 
 \begin{table*}
 	\setlength{\tabcolsep}{1.15mm}  
 	\begin{tiny}
 		\caption{Line-by-line abundance information for every possible member of NGC 6522.}
 		\begin{tabular}{lcccccc}
 			\hline
 			Element  &  $\lambda^{air} ({\rm \AA{}})$ & 2M18032356$-$3001588  & 2M18034052$-$3003281  &  2M18033965$-$3000521 & 2M18033819$-$3000515  & 2M18033660$-$3002164 \\
 			\hline
 			{ Fe I} &15194.492&  6.23    &   -    &  6.43   & 6.49 &     -     \\
 			&15207.526&  6.15    & 6.36 &  6.50   & 6.51  &  6.17   \\
 			&15395.718&  6.35    & 6.49 &  6.52   &   -     &  6.38   \\
 			&15490.339&  6.24    & 6.42 &  6.52   & 6.58  &    -      \\
 			&15648.510&  6.23    & 6.38 &  6.46   & 6.50  &    -      \\
 			&15964.867&  6.31    & 6.57 &  6.44   & 6.42  &   6.49  \\
 			&16040.657&  6.20    & 6.54 &  6.39   & 6.42  &   6.38  \\
 			&16153.247&  6.25    & 6.50 &  6.45   & 6.50  &   6.46  \\
 			&16165.032&  6.23    & 6.57 &  6.42   & 6.43  &   6.35   \\
 			${\rm \langle A(Fe) \rangle }\pm \sigma$ & & 6.25$\pm$0.05 &  6.48$\pm$0.08  & 6.46$\pm$0.04 & 6.48$\pm$0.05 & 6.37$\pm$0.10 \\
 			\hline
 			{ Al I}   & 16719.0  &  5.54   &  6.41    &  5.75   &  6.16     &     5.68   \\
 			& 16750.0  &  5.56   &  6.32    &  5.78   &  6.09     &      -       \\
 			& 16763.0  &    -      &  6.26    &     -     &  6.26     &      -       \\
 			${\rm \langle A(Al) \rangle }\pm \sigma$ & & 5.55$\pm$0.01 & 6.33$\pm$0.06  & 5.77$\pm$0.02 & 6.17$\pm$0.07 & 5.68 \\
 			\hline
 			{ Mg I} & 15740.7  &  6.53   &   6.63  & 6.58     & 6.64    &   6.99    \\
 			& 15748.9  &  6.52   &   6.68  & 6.59     & 6.68    &   6.72   \\
 			& 15765.8  &  6.41   &   6.53  &   -     & 6.49    &   6.63   \\                
 			${\rm \langle A(Mg) \rangle }\pm \sigma$ & & 6.49$\pm$0.05 & 6.61$\pm$0.06 & 6.59$\pm$0.05 &  6.60$\pm$0.08  &  6.78$\pm$0.15 \\
 			\hline
 			{ Si I}   &  15361.1  &  -           &   -          &     -     &    -       &     -     \\
 			&  15376.8  &  -           &   -          &     -     &    -       &     -      \\
 			&  15557.8  &  6.62      &   6.86     &     -     &   6.68   &     -       \\
 			&  15884.5  &  6.39      &   6.76     &   6.63  &   6.63   &     -       \\
 			&  15960.1  &  6.53      &   7.10     &   6.65  &   6.67   &    6.82   \\
 			&  16060.0  &  6.64      &   6.97     &   6.82  &   6.71   &     -       \\
 			&  16094.8  &  6.61      &   6.92     &   6.72  &   6.79   &    6.66   \\
 			&  16129.0  &     -        &     -        &     -     &     -      &     -       \\
 			&  16163.7  &  6.71      &   7.00     &     -     &   6.79   &     -       \\
 			&  16170.2  &    -         &     -        &     -     &     -      &     -       \\
 			&  16215.7  &  6.53      &   6.92     &   6.70  &   6.68   &     -       \\
 			&  16241.8  &    -         &   6.83     &   6.65  &   6.63   &     -       \\
 			&  16680.8  &  6.56      &   6.88     &   6.74  &   6.64   &    7.33  \\
 			&  16828.2  &  6.74      &   6.89     &     -     &   6.89   &      - \\                                                                                 
 			${\rm \langle A(Si) \rangle }\pm \sigma$ &  & 6.59$\pm$0.10  &   6.91$\pm$0.09   &     6.70$\pm$0.06    &     6.71$\pm$0.08  & 6.94$\pm$0.29    \\
 			\hline
 			{ Ce II} & 15277.65 &   -        &    -     &    -      &    -       &  -    \\
 			& 15784.75 &   0.48   &    -     &    -      &    -       &  -     \\
 			& 15958.40 &   0.50   &    -     &    -      &  0.88    &  -     \\
 			& 15977.12 &    -       &    -     &    -      &  0.83    &  -    \\
 			& 16327.32 &    -       &    -     &    -      &    -       &  -    \\	                
 			& 16376.48 &   0.40   &    -     &    -      &  0.81    &  -    \\
 			& 16595.18 &   0.50   &    -     &    -      &  0.84    &  -    \\
 			& 16722.51 &    -       &    -     &    -      &    -       &  -    \\
 			${\rm \langle A(Ce) \rangle }\pm \sigma$ & & 0.47$\pm$0.04 &    -   &    -     &     0.84$\pm$0.03    &  -  \\
 			\hline
 			{ $^{12}$C from $^{12}$C$^{16}$O lines} 	& 15774 $-$ 15787  &   6.67                  &  -                        &  -      &  7.10  &  - \\
 			& 15976 $-$ 16000 &   6.75                  &  7.20                   & 7.15  &  7.13 &   - \\
 			&  16183 $-$ 16196 &   6.71                  &  7.16                   & 7.07  &  7.04 &   - \\
 			${\rm \langle A(C) \rangle }\pm \sigma$       &                               &  6.71$\pm$0.03 &  7.18$\pm$0.02 &  7.11$\pm$0.04 &  7.09$\pm$0.04 & -\\
 			\hline
 			{ $^{14}$N from $^{12}$C$^{14}$N lines}	&15260.&    7.88    &  8.01  &   -     &  8.14  &   -  \\
 			&15322.&    7.97    &  8.08  &  7.83 &  8.18  &   -  \\
 			&15397.&      -       &    -     &   -     &  8.06  &   -  \\
 			&15332.&   7.86    &  8.10  &  7.88 &  8.10  &  7.70   \\
 			&15410.&   7.88    &  8.11  &  7.87 &  8.13  &    - \\
 			&15447.&   7.82    &  7.80  &    -    &  8.09  &    - \\
 			&15466.&   7.78    &  7.89  &  7.78 &  8.05  &  7.95   \\
 			&15472.&   7.92    &  8.04  &    -    &  8.17  &    -  \\
 			&15482.&   7.81    &  8.04  &  7.76 &  8.06  &    -  \\
 			${\rm \langle A(N) \rangle }\pm \sigma$       &           &   7.87$\pm$0.06  & 8.01$\pm$0.10 & 7.82$\pm$0.05  & 8.11$\pm$0.05 & 7.83$\pm$0.13 \\
 			\hline
 			{ $^{16}$O from $^{16}$OH lines}        & 15278.524  &  7.93  &   -     &   -     &  7.87    & -  \\
 			&  15281.052 &  7.86  &  8.02 &   -     &  8.10    & - \\
 			&  15390.8     &  7.81  &    -    &  7.97 &  8.02    & - \\
 			&  15568.780 &  7.95  &  8.06 &   -     &  8.04    & - \\
 			&  16190.132 &  7.76  &    -    &  7.94 &  7.94    & - \\
 			& 16192.130  &  7.80  &    -    &   -     &  7.98    & - \\    
 			${\rm \langle A(O) \rangle }\pm \sigma$ &                    & 7.85$\pm$0.07 & 8.04$\pm$0.02  &   7.96$\pm$0.02  &   7.99$\pm$0.07 & - \\
 			\hline
 		\end{tabular}  \label{table3}
 	\end{tiny}
 \end{table*}

\end{appendix}


\end{document}